\documentclass[sn-mathphys-num, iicol]{sn-jnl}

\usepackage{graphicx}%
\usepackage{multirow}%
\usepackage{amsmath,amssymb,amsfonts}%
\usepackage{amsthm}%
\usepackage{mathrsfs}%
\usepackage[title]{appendix}%
\usepackage{xcolor}%
\usepackage{textcomp}%
\usepackage{manyfoot}%
\usepackage{booktabs}%
\usepackage{algorithm}%
\usepackage{algorithmicx}%
\usepackage{algpseudocode}%
\usepackage{listings}%

\usepackage{stfloats}
\usepackage{caption}
\usepackage{subcaption}
\usepackage{tikz}
\usepackage{comment}
\DeclareMathOperator*{\argmin}{arg\,min}

\newcommand{\vv}{\bold{v}}

\newcommand{\R}{\mathbb{R}}
\newcommand{\dint}{\displaystyle\int}
\newcommand{\dsum}{\displaystyle\sum}

\theoremstyle{thmstyleone}%
\theoremstyle{thmstyletwo}%

\theoremstyle{thmstylethree}%

\begin{document}

\title[Enhancing Dynamic CT Image Reconstruction with Neural Fields and Optical Flow]{Enhancing Dynamic CT Image Reconstruction with Neural Fields and Optical Flow}


\author*[1]{\fnm{Pablo} \sur{Arratia}}\email{pial20@bath.ac.uk}

\author[1]{\fnm{Matthias J.} \sur{Ehrhardt}}\email{me549@bath.ac.uk}

\author[1]{\fnm{Lisa} \sur{Kreusser}}\email{lmk54@bath.ac.uk}

\affil*[1]{\orgdiv{Department of Mathematical Sciences}, \orgname{University of Bath}, \orgaddress{\city{Bath}, \postcode{BA2 7AY}, \country{UK}}}


\abstract{In this paper, we investigate image reconstruction for dynamic Computed Tomography. The motion of the target with respect to the measurement acquisition rate leads to highly resolved in time but highly undersampled in space measurements. Such problems pose a major challenge: not accounting for the dynamics of the process leads to a poor reconstruction with non-realistic motion. Variational approaches that penalize time evolution have been proposed to relate subsequent frames and improve image quality based on classical grid-based discretizations. Neural fields have emerged as a novel way to parameterize the quantity of interest using a neural network with a low-dimensional input, benefiting from being lightweight, continuous, and biased towards smooth representations. The latter property has been exploited when solving dynamic inverse problems with neural fields by minimizing a data-fidelity term only. We investigate and show the benefits of introducing explicit   motion regularizers for dynamic inverse problems based on partial differential equations, namely, the optical flow equation, for the optimization of neural fields. We compare it against its unregularized counterpart and show the improvements in the reconstruction. We also compare neural fields against a grid-based solver and show that the former outperforms the latter in terms of PSNR in this task.}




\maketitle

\section{Introduction}\label{sec1}

In many imaging tasks, the target object changes during the data acquisition. In clinical settings for instance, imaging techniques such as Computed Tomography (CT), Positron Emission Tomography (PET) or Magnetic Resonance Imaging (MRI) are used to study moving organs such as the heart or the lungs. Usually, the acquired data is a time series collected at several finely discretized times $0=t_1<\ldots<t_{N_T}=T$. However, the motion of these organs prevents the scanners from taking enough measurements at a single time instance, resulting in highly undersampled spatial measurements. A naive way to proceed is by neglecting the time component and solving several static inverse problems. However, the lack of information makes this frame-by-frame reconstruction a severely ill-posed problem leading to a poor reconstruction. A common procedure is to bin the data in time, where several time-step measurements are collapsed into one to gain more information in space at the cost of losing temporal resolution and introducing artefacts in the reconstruction. It is therefore necessary to seek a spatiotemporal quantity with coherence between subsequent frames whose reconstruction considers the dynamics of the process. 

\subsection{Dynamic Inverse Problems}

In (static) inverse problems, we aim to reconstruct a quantity $u:\Omega\subset\R^d\to\R$ from a discrete set of measurements $f$ obtained by some device by solving an equation of the form
\begin{equation}\label{eq:static inverse problem}
Ku+\varepsilon = f.
\end{equation}
Here $K$ is the forward operator that models the imaging process by mapping the continuous object $u$ into a discrete set of measurements, and $\varepsilon$ is the noise coming from the measurement acquisition. 

For dynamic inverse problems, we formulate the problem as follows: we let $\bold{f}=\{f_{t_1},\ldots,f_{t_{N_T}}\}\subset\R^M$ be the measurements at several time steps, with $M$ the number of measurements collected at a given time. The goal is to recover a time-dependent quantity $u:\Omega_T:=\Omega\times[0,T]\to\R$ by solving the equation below 
\begin{equation}\label{eq:dip}
    K_t[u_t]+\varepsilon_t = f_t, \quad\text{ for } t\in\{t_1,\ldots,t_{N_T}\}.
\end{equation} 
Here $u_t$, $K_t[u_t]\in\R^M$, and $\varepsilon_t$ are the solution, the imaging process, and the noise at time $t$, respectively.

The classical way to address this problem is to discretize the solution using a grid-based representation $u\in\R^{N\times N_T}$, the Casorati matrix, with $N$ the number of pixels in space and $N_T$ the number of frames. The columns of this matrix represent the solution at the corresponding time step. The problem is then solved using a variational formulation which consists of a data-fidelity term plus some suitable regularizer $\mathcal{R}$ that seeks correlation between the columns of $u$. Common examples of such regularizers include sparsity-based regularizers, inspired by the idea that the sought quantity can be compactly represented on a suitable basis, e.g., total variation, wavelets or shearlets, see \cite{rudin1992nonlinear, colonna2010radon, bubba2017shearlet, bubba2023efficient}; the nuclear norm, which promotes the solution to be a low-rank matrix, see \cite{lingala2011accelerated}; first-order time derivative penalizers, suitable for sequences with small displacements, see \cite{steeden2018real, niemi2015dynamic}. 

Another option is explicitly considering the target's motion by including its velocity field $\vv$ in the formulation. We call this a motion-based regularizer, where a partial differential equation (PDE) $r(u,\vv)=0$, relating $u$, the velocity field $\vv$, and their derivatives, is used to impose a physical prior. In this paper, we set the motion model as the so-called \emph{Optical Flow equation}:
\begin{equation}\label{eq:optical flow}
r(u,\vv) := \partial_t u + \vv\cdot\nabla u = 0, \quad \text{ in } \Omega_T.
\end{equation}  
This equation shall be discussed in more detail in section \ref{sec:motion models}. As the underlying motion is unknown the overall problem is commonly referred to as a joint image reconstruction and motion estimation task, see \cite{Burger_Dirks_Schonlieb_2018, burger2015nonlinear, burger2017variational, aviles2021compressed}. We refer to \cite{hauptmann2021image} for an extensive review of dynamic inverse problems. 

Classical grid-based representations of the spatiotemporal image suffer from two issues: (1) their lack of regularity which motivates the use of several regularizers such as the ones mentioned above, and (2) their complexity grows exponentially with the dimension and polynomially with the discretization due to the curse of dimensionality which can incur in memory burden. The latter is particularly relevant to large-scale problems, for instance, when the measurement frame rate and/or spatial resolution is high, in 3D+time domains, etc. In the next section, we introduce neural fields, an alternative continuous representation using deep neural networks.

\subsection{Neural Fields}

In recent years, coordinate-based multilayer perceptrons (MLPs) have been employed as a new way of parameterizing quantities of interest. In computer vision, these are referred to as neural fields or implicit neural representations \cite{neural_fields, SIREN}, while the term Physics-Informed Neural Networks (PINNs) has been adopted when used to solve PDEs \cite{pinn, pinnsurvey}. The main idea is to use a neural network $u_{\theta}$ with trainable weights $\theta$ as an ansatz for the solution of the problem. It takes as input a spatio-temporal point $(x,t)\in\Omega_T$, and outputs the value $u_{\theta}(x,t)$ at that point, e.g., the intensity of the image at that particular time and location. The problem is then rephrased as a non-convex optimization that seeks optimal weights $\theta$. The method requires training a neural network for every new instance, thus, it is said to be self-supervised and differs from the usual learning framework where a solution map is found by training a network over large datasets. Applications of neural fields include image reconstruction in CT \cite{zang2021intratomo, sun2021coil, reed2021dynamic}, MRI \cite{xu2023nesvor, kunz2024implicit, huang2023neural, feng2022spatiotemporal, catalan2023unsupervised}, image registration \cite{wolterink2022implicit, lopez2023warppinn, zou2023homeomorphic}, continuous shape representation via signed distance functions \cite{alblas2022going}, view synthesis with Neural Radiance Fields (NeRF) \cite{mildenhall2021nerf}, among others.

It is well-known that, under mild conditions, neural networks can approximate functions at any desired tolerance \cite{Hornik_Stinchcombe_White_1989}, but other properties have justified their widespread use:
\begin{enumerate}
\item Implicit regularization. Numerical experiments and theoretical results show that neural fields tend to learn smooth functions early during training, commonly referred to as spectral bias \cite{rahaman2019spectral, jacot2018neural, wang2021eigenvector}. This is both advantageous and disadvantageous: neural fields can capture smooth regions of natural images but will struggle at capturing sharp edges. The latter can be overcome with Fourier feature encoding \cite{tancik2020fourier} or with sinusoidal activation functions as in SIREN \cite{SIREN}. We highlight that smoothness in time is highly desirable for dynamic inverse problems.
\item Overcoming the curse of dimensionality. In \cite{hutzenthaler2020proof, jentzen2018proof} it is shown that the amount of weights needed to approximate the solution of particular PDEs grows polynomially on the dimension of the domain. For the same reason, only a few weights can represent complex images, leading to a lightweight, compact and memory-efficient representation. This has been exploited, for instance, in \cite{martel2021acorn} for image compression.
\item Continuous and differentiable representation. This property is exploited in PINNs where the derivatives of a PDE are computed using automatic differentiation with an accuracy only limited by machine precision. In particular, the parametrization does not rely on a mesh as in finite differences or finite elements.
\end{enumerate}

In the context of dynamic inverse problems and neural fields, part of the literature relies entirely on the smoothness introduced by the network on the spatial and temporal variables to get a regularized solution. This motivates the minimization of the data-fidelity term without considering any explicit regularizers. Examples can be found on dynamic cardiac MRI in \cite{kunz2024implicit, catalan2023unsupervised}, where the network outputs the real and imaginary parts of the signal. In \cite{ zang2021intratomo, sun2021coil, wu2023self} a neural field is used to inpaint the undersampled sinogram. Once optimized, inference is performed by rendering the network at a regular grid and applying a suitable transformation, e.g., filtered back projection. Regularized variational problems with neural fields have been considered in \cite{huang2023neural, liu2022recovery} with a regularization-by-denoising approach for MRI and intensity diffraction tomography respectively, while \cite{xu2023nesvor, feng2022spatiotemporal} approximate a total variation regularizer with finite differences. Such approaches do not exploit the continuous and differentiable representation offered by neural fields. In \cite{lozenski2022memory, lozenski2024proxnf} neural fields are used to solve a photoacoustic tomography dynamic reconstruction emphasizing their memory efficiency and using total variation and Tikhonov regularization respectively with gradients computed via automatic differentiation. Dynamic CT has been addressed in 3D+time domains in \cite{reed2021dynamic, zhang2023dynamic}, where the neural field parametrizes the initial frame and a deformation vector field warps it to get the subsequent frames in time. A similar idea is used for novel view synthesis for dynamic scenes in D-NeRF \cite{pumarola2021d}. 

\subsection{Contributions}

Motivated by \cite{Burger_Dirks_Schonlieb_2018, burger2017variational}, in this paper, we study neural fields in the context of dynamic inverse problems in a highly undersampled measurement regime with the optical flow equation as an explicit PDE-based motion regularizer imposed as a soft constraint as in PINNs. We leverage the arbitrary resolution and automatic differentiation of neural fields to compute spatial and temporal derivatives. We do not consider the nuclear norm and the sparsity-based regularizers previously mentioned since they act on a discrete representation of the solution on a cartesian grid and hence do not exploit the mesh-free nature of the neural field: to use them, the neural field needs to be queried at points on a cartesian grid to get a discrete representation over which the regularizer can act.

Our findings are based on numerical experiments on dynamic CT performed on three synthetic and one real datasets, all in a 2D+time domain. The contributions are summarized as follows:
\begin{itemize}
    \item Constraining the neural field with an explicit motion model ensures that only physically feasible solution manifolds are considered. We demonstrate that our approach improves the reconstruction when compared to a motionless unregularized model.
    \item We show how to leverage the mesh-free nature of neural fields to impose regularizers for imaging tasks. 
    
    \item We study the reconstruction obtained by neural fields and by a grid-based representation. We demonstrate that neural fields can outperform classical discretizations in terms of the quality of the reconstruction for highly undersampled dynamic CT.
    
\end{itemize}

The paper is organized as follows: in section \ref{sec:dynamic inverse problems} we introduce dynamic compute tomography, motion models with the optical flow equation, and the joint image reconstruction and motion estimation variational problem as in \cite{Burger_Dirks_Schonlieb_2018}; in section \ref{sec:methods} we state the main variational problem to be minimized and study how to solve it with neural fields and with a grid-based representation; in section \ref{sec:datasets} we describe the datasets; in section \ref{sec:numerical experiments} we investigate our method numerically and show its improvements in comparison to unregularized neural fields and the grid-based representation; we finish with the conclusions in section \ref{sec:conclusions}.

\section{Dynamic Inverse Problems for Computed Tomography}\label{sec:dynamic inverse problems}

In X-ray CT \cite{radon20051, smith1977practical, natterer2001mathematics} the unknown $u$ is the non-negative absorption of photons of the imaged object. The forward model is a line integral whose precise formulation depends on the scanner: common geometries include parallel and fan beams. We focus on the latter, where the X-rays are emitted from several source points $p\in\{p_1,\ldots,p_P\}\subset\R^d$ through the object $O$ in different directions towards $M$ sensors that collect the photons, see figure \ref{fig:fan beam}. Given a source point $p\in\R^d$, a set $\{L^p_1,\ldots,L^p_M\}$ of $M$ lines is constructed with $L^p_j$ going from $p$ to the $j$-th sensor. The projection from $p$ can be represented as follows:
\[K[u](p) = \left[\dint_{L^p_1}u(x)dx,\ldots,\dint_{L^p_M}u(x)dx\right]^T,\]
leading to $f\in\R^{M\times P}$ measurements when considering all source positions $p_1,\ldots, p_P$.

\begin{figure}[]
    \centering
    \includegraphics[width=1\linewidth]{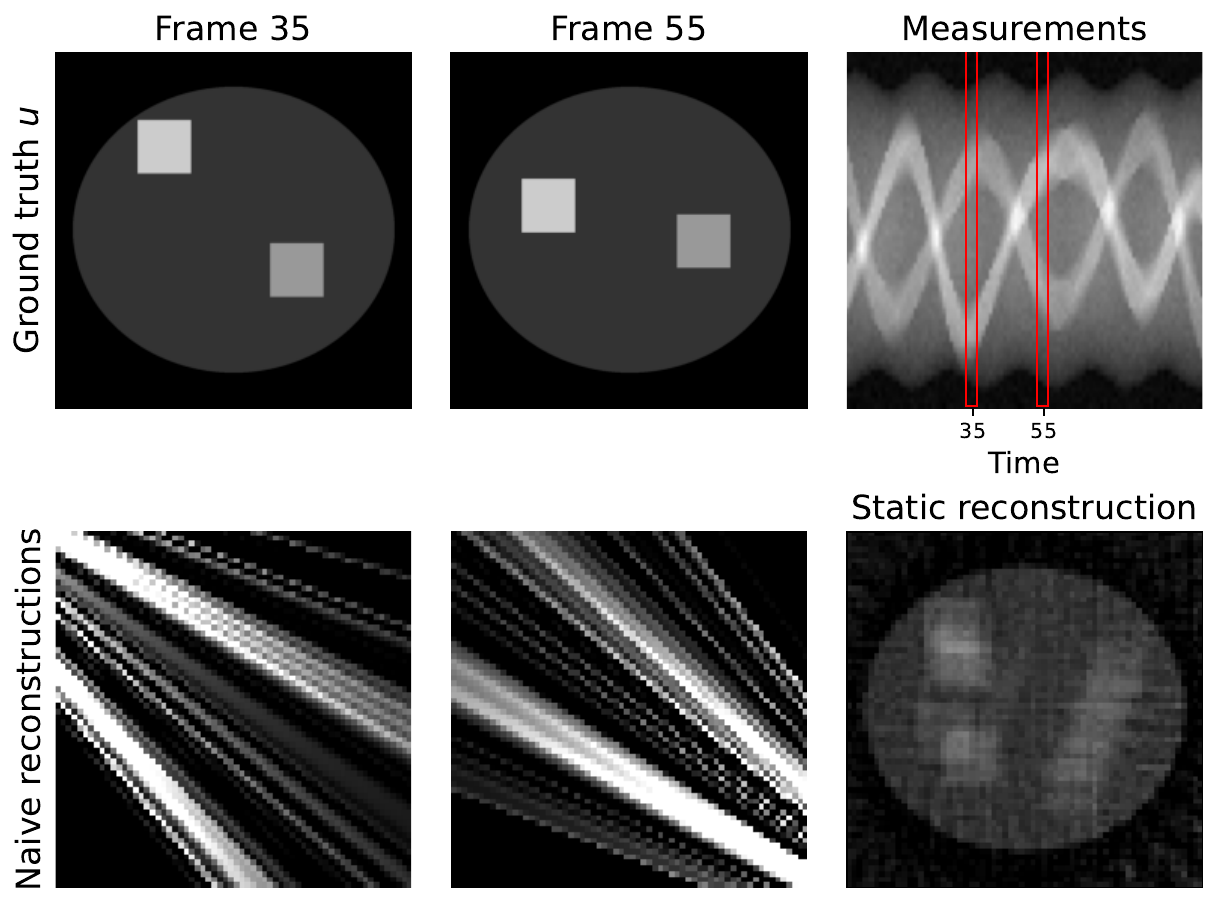}
    \caption{Top row: ground truth image $u$ at frames 35 and 55 (out of 100) with two squares moving, and measurements $\mathbf{f}\in\R^{M\times N_T}$. The red boxes indicate the projection obtained in frames 35 and 55. Bottom row: naive reconstructions using filtered back projection. The first two figures represent a frame-by-frame solution from measurements at frames 35 and 55. The third figure represents the static reconstruction from $f_{\text{bin}}$.}
    \label{fig:two-square naive}
\end{figure}

For dynamic CT we consider both the object's motion and the scanner's rotation around it. Hence, the forward operator $K_t$ in \eqref{eq:dip} is the projection from a source point $p_t$:
\[K_t[u_t]:=K[u_t](p_t).\]
When the motion of $u$ is slow compared to the rotation of the scanner, it is possible to pose equation \eqref{eq:dip} as \eqref{eq:static inverse problem} by neglecting the time variable. In this case, $u$ can be reconstructed from the binned vector $f_{\text{bin}}:=[f_{t_1},\ldots,f_{t_{N_T}}]\in\R^{M\times N_T}$. To highlight the necessity of motion models, two naive reconstructions obtained using filtered back projection are shown in figure \ref{fig:two-square naive}. The first row depicts the two-square phantom we will introduce in section \ref{sec:datasets} for our numerical experiments, and the measurements. The reconstructions are shown at the bottom. As expected, we cannot get a reliable reconstruction from one projection only. The image on the bottom right corresponds to the reconstruction from $f_{\text{bin}}$. The result is an image that blurs the motion of the squares.

In the next section, we introduce motion models to regularize the dynamic inverse problem.

\begin{figure}
    \centering
    \begin{tikzpicture}[x=0.75pt,y=0.75pt,yscale=-0.8,xscale=0.8]

\draw  (139.5,156.6) -- (472.5,156.6)(309.5,25) -- (309.5,261.6) (465.5,151.6) -- (472.5,156.6) -- (465.5,161.6) (304.5,32) -- (309.5,25) -- (314.5,32)  ;
\draw  [color={rgb, 255:red, 0; green, 0; blue, 0 }  ,draw opacity=1 ][fill={rgb, 255:red, 0; green, 0; blue, 0 }  ,fill opacity=1 ] (316,156.6) .. controls (316,153.01) and (313.09,150.1) .. (309.5,150.1) .. controls (305.91,150.1) and (303,153.01) .. (303,156.6) .. controls (303,160.19) and (305.91,163.1) .. (309.5,163.1) .. controls (313.09,163.1) and (316,160.19) .. (316,156.6) -- cycle ;
\draw   (266,123) .. controls (286,113) and (339,95) .. (365,113.6) .. controls (391,132.2) and (359,140) .. (363,180.6) .. controls (367,221.2) and (286,213) .. (266,183) .. controls (246,153) and (246,133) .. (266,123) -- cycle ;
\draw  [color={rgb, 255:red, 208; green, 2; blue, 27 }  ,draw opacity=1 ][fill={rgb, 255:red, 208; green, 2; blue, 27 }  ,fill opacity=1 ] (415.5,125.1) .. controls (415.5,121.51) and (412.59,118.6) .. (409,118.6) .. controls (405.41,118.6) and (402.5,121.51) .. (402.5,125.1) .. controls (402.5,128.69) and (405.41,131.6) .. (409,131.6) .. controls (412.59,131.6) and (415.5,128.69) .. (415.5,125.1) -- cycle ;
\draw  [dash pattern={on 4.5pt off 4.5pt}]  (406.13,125.98) -- (159.74,201.18) ;
\draw [shift={(409,125.1)}, rotate = 163.03] [fill={rgb, 255:red, 0; green, 0; blue, 0 }  ][line width=0.08]  [draw opacity=0] (8.93,-4.29) -- (0,0) -- (8.93,4.29) -- cycle    ;
\draw  [color={rgb, 255:red, 208; green, 2; blue, 27 }  ,draw opacity=1 ][fill={rgb, 255:red, 208; green, 2; blue, 27 }  ,fill opacity=1 ] (429.5,176.6) .. controls (429.5,173.01) and (426.59,170.1) .. (423,170.1) .. controls (419.41,170.1) and (416.5,173.01) .. (416.5,176.6) .. controls (416.5,180.19) and (419.41,183.1) .. (423,183.1) .. controls (426.59,183.1) and (429.5,180.19) .. (429.5,176.6) -- cycle ;
\draw  [dash pattern={on 4.5pt off 4.5pt}]  (420.01,176.88) -- (159.74,201.18) ;
\draw [shift={(423,176.6)}, rotate = 174.67] [fill={rgb, 255:red, 0; green, 0; blue, 0 }  ][line width=0.08]  [draw opacity=0] (8.93,-4.29) -- (0,0) -- (8.93,4.29) -- cycle    ;
\draw   (159.74,201.18) -- (380.49,23.32) -- (440.56,239.95) -- cycle ;
\draw  [color={rgb, 255:red, 248; green, 231; blue, 28 }  ,draw opacity=1 ][fill={rgb, 255:red, 248; green, 231; blue, 28 }  ,fill opacity=1 ] (166.24,201.18) .. controls (166.24,197.59) and (163.33,194.68) .. (159.74,194.68) .. controls (156.15,194.68) and (153.24,197.59) .. (153.24,201.18) .. controls (153.24,204.77) and (156.15,207.68) .. (159.74,207.68) .. controls (163.33,207.68) and (166.24,204.77) .. (166.24,201.18) -- cycle ;

\draw (346,141.4) node [anchor=north west][inner sep=0.75pt]    {$\theta $};
\draw (150,216.4) node [anchor=north west][inner sep=0.75pt]    {$p$};
\draw (289.5,136) node [anchor=north west][inner sep=0.75pt]    {$O$};
\draw (408,62) node [anchor=north west][inner sep=0.75pt]   [align=left] {Detector};
\draw (372,182.4) node [anchor=north west][inner sep=0.75pt]    {$L_{i}^{p}$};
\draw (377,130.4) node [anchor=north west][inner sep=0.75pt]    {$L_{j}^{p}$};
\draw (417,108.4) node [anchor=north west][inner sep=0.75pt]    {$f_{p,j}$};
\draw (435,169.4) node [anchor=north west][inner sep=0.75pt]    {$f_{p,i}$};

\end{tikzpicture}

    \caption{Fan-beam geometry for CT. X-rays emitted from the source point $p$ in different directions are attenuated by the object and then measured in the sensors represented by the red dots.}
    \label{fig:fan beam}
\end{figure}

\subsection{Motion Model}\label{sec:motion models}

A motion model describes the relation between pixel intensities $u$ and the velocity flow $\vv$ through an equation $r(u,\vv)=0$ in $\Omega_T$. See, e.g., equation \eqref{eq:optical flow}. Its choice is application-dependent, for instance, the continuity equation imposes a mass-preservation constraint, while the optical flow equation promotes no change in the intensities. These models are typically employed for the task of motion estimation, this is, determining the velocity flow $\vv$ for the given image sequence $u$.

In this work, we focus on the well-known optical flow equation introduced in \eqref{eq:optical flow}. It is derived from the brightness constancy assumption, which states that pixels keep constant intensity along their trajectory in time. This model poses a scalar equation for the $d$ components of the velocity field, leading to an underdetermined equation. This can be solved by considering a variational problem in $\vv$ with a regularization term:
\begin{equation}\label{eq:variational problem v}
\min_{\vv}\mathcal{A}(r(u,\vv))+\beta\mathcal{S}(\vv),
\end{equation}
where $\mathcal{A}$ is a metric measuring how well the equation $r(u,\vv)=0$ is satisfied, $\mathcal{S}$ is a regularizer, and $\beta>0$ is the regularization parameter balancing both terms. This variational model was firstly introduced in \cite{Horn_Schunck_1981} with $\mathcal{A}$ as the $L^2$-norm and $\mathcal{S}$ as the $L^2$-norm of the gradient. Since then, different norms and regularizers have been tried, for instance, in \cite{aubert1999computing} the $L^1$-norm is used to impose the motion model, same as in \cite{zach2007duality} which employs the total variation for regularization.

\subsection{Joint Image Reconstruction and Motion Estimation}

To solve highly undersampled dynamic inverse problems, a joint variational problem is proposed in \cite{Burger_Dirks_Schonlieb_2018} where not only the dynamic process $u$ is sought, but also the underlying motion expressed in terms of a velocity field $\vv$. The main hypothesis is that a joint reconstruction can enhance the discovery of both quantities, image sequence and motion, improving the final reconstruction compared to motionless models. Hence, the sought solution is a minimizer for the variational problem:
\begin{equation}\label{eq:variational problem}
\begin{aligned}
\min_{u,\vv}&\mathcal{D}(u,\bold{f}) + \alpha\mathcal{R}(u) + \beta \mathcal{S}(\vv) + \gamma \mathcal{A}(r(u,\vv)),
\end{aligned}
\end{equation}
where $\alpha,\beta,\gamma>0$ are regularization parameters balancing the four terms. We also recall $\bold{f} = \{f_{t_1},\ldots,f_{t_{N_T}}\}$. In \cite{Burger_Dirks_Schonlieb_2018}, it is shown,  among other things, how the pure motion estimation task of a noisy sequence can be enhanced by solving the joint task of image denoising and motion estimation.

This model was further employed for 2D+time problems in \cite{burger2017variational} and \cite{aviles2021compressed}. In the former, its application on dynamic CT is studied with sparse limited angles using both the $L^1$ and $L^2$-norms for the data fidelity term, with better results for the $L^1$-norm. In the latter, the same logic is used for dynamic cardiac MRI. In 3D+time domains, we mention \cite{djurabekova2019application} and \cite{lucka2018enhancing} for dynamic CT and dynamic photoacoustic tomography respectively. More methods for dynamic CT have been proposed based on the simultaneous algebraic reconstruction technique and motion compensation. We refer to the interested reader to \cite{wang2013simultaneous, chee2019mcsart}

\section{Methods}
\label{sec:methods}

Different data-fidelity terms can be considered depending on the nature of the noise. In this work, we consider Gaussian noise $\varepsilon$. To satisfy equation \eqref{eq:dip} at time $t$ we use an $L^2$ distance between predicted measurement and data $f_t$:
\[\mathcal{D}_t(u, f_t) := \dfrac{1}{2}\|K_{t}[u_{t}]-f_t\|^2_{2}.\]
The overall data-fidelity term in \eqref{eq:variational problem} is a mean over the measured times:
\begin{equation}\label{eq:data fidelity}
\mathcal{D}(u,\bold{f}):=\dfrac{1}{N_T}\dsum_{i=1}^{N_T} \mathcal{D}_{t_i}(u, f_{t_i}).
\end{equation}
Since $u$ represents a natural image, a suitable choice for the regularizer $\mathcal{R}$ is the total variation in space to promote noiseless images and capture edges. For the motion model, we consider the optical flow equation \eqref{eq:optical flow}, and to measure its discrepancy to 0 we use the $L^1$-norm; for the regularizer in $\vv=(v_1,\ldots, v_d)^T$ we consider the total variation on each of its components:
\begin{equation}\label{eq:regularizers}
\begin{aligned}
\mathcal{R}(u)&:=\dint_{\Omega_T}\|\nabla u\|_2,\\
\mathcal{A}(r(u,\vv)) &:=\dint_{\Omega_T}|\partial_tu+ \vv\cdot\nabla u|,\\
\mathcal{S}(\vv)&:=\dint_{\Omega_T}\dsum_{j=1}^d\|\nabla v_j\|_2.
\end{aligned}
\end{equation}
For conciseness, we have omitted the dependency of the integrand on $(x,t)$.

We now describe how to solve the variational problem \eqref{eq:variational problem} numerically with neural fields. We proceed with a discretize-then-optimize approach. We also provide a brief description of the grid-based approach in \cite{Burger_Dirks_Schonlieb_2018} as we will compare it against our method.

\subsection{Numerical evaluation with Neural Fields}\label{sec:neural field method}

We parametrize both the image and the motion with two independent neural fields. Both take a point $(x,t)\in\R^3$ as input, then a Fourier feature embedding \cite{tancik2020fourier} is applied independently on the space and time variables, and then apply $L+1$ fully-connected transformations. This is described below:
\begin{equation}\label{eq:architecture}
\begin{aligned}
\boldsymbol{x}_0 &= (\Gamma_1(x),\Gamma_2(t)) \in \R^{m} \\
\boldsymbol{x}_l &= \sigma( W^l\boldsymbol{x}_{l-1}+b^l) \in\R^{d_l}, \quad l=1,\ldots, L,\\
\boldsymbol{x}_{L+1} &=  W^{L+1}\boldsymbol{x}_{L}+b^{L+1}\in\R^{d_{L+1}},
\end{aligned}
\end{equation}
where $\{(W^l,b^l)\}_{l=1}^{L+1}$ are the weights and biases and $\sigma:\R\to\R$ is the non-linear activation function acting element-wise. The Fourier embeddings are defined as $\Gamma_1(x):= (\sin(2\pi \mathbf{B_x} x), \cos(2\pi \mathbf{B_x} x))\in\R^{2m_x}$ and $\Gamma_2(t):= (\sin(2\pi \mathbf{B_t} t), \cos(2\pi \mathbf{B_t} t))\in\R^{2m_t}$, with the sinusoidal functions acting element-wise. The matrices $\mathbf{B_x}\in\R^{m_x\times 2}$ and $\mathbf{B_t}\in\R^{m_t\times 1}$ have non-trainable entries sampled from gaussian distributions $(\mathbf{B_x})_{ij}\sim \mathcal{N}(0,\sigma_x^2)$ and $(\mathbf{B_t})_{ij}\sim \mathcal{N}(0,\sigma_t^2)$. Here $\sigma_x$ and $\sigma_t$ are hyperparameters accounting for the frequencies the neural field can capture; the larger they are, the more frequencies can be captured earlier during optimization.
 
We let $u_{\theta}$ and $\vv_{\phi}$ to be the image and the velocity field, respectively, with $\theta$ and $\phi$ denoting their weights and biases. Clearly, we set $d_{L+1}=1$ for $u_{\theta}$ and $d_{L+1}=2$ for $\vv_{\phi}$. Even though $u_{\theta}$ is a composition of known functions, there is no closed-form expression for its X-ray transform $K_t[(u_{\theta})_{t}]$. For this reason, the numerical evaluation of the forward operator is performed by first evaluating the network at points on a cartesian grid to get a grid-based representation at time $t$ as $(u_{\theta})_{t}:=\{u_{\theta}(x_j, t)\}_{j=1,\ldots,N}$, and then applying our preferred implementation of the X-ray transform. In particular, we work with \emph{Tomosipo} \cite{hendriksen-2021-tomos}, a library that provides an integration of the \emph{ASTRA-toolbox} \cite{van2015astra, van2016fast} with PyTorch, making it suitable for the optimization of neural networks. The data fidelity term in \eqref{eq:data fidelity} requires first the evaluation of the network at $N\times N_T$ fixed grid-points to get the scene $\{u_{\theta}(x_j,t_i)\}_{i=1,\ldots,N_T;j=1,\ldots,N}$, and, second, the application of the forward models $\{K_{t_i}\}_{i=1\ldots,N_T}$ to each frame. We call this the full-batch approach. This might be expensive and time-consuming. A common practice is to proceed with a mini-batch-like approach in time to speed up the optimization and avoid poor local minima. In this setting, at each iteration, we randomly sample $1\leq N_B\leq N_T$ frames, say, $\{i_1,\ldots,i_{N_B}\}$, and the neural field is evaluated at the corresponding points to get the representation of the image at times $\{t_{i_1},\ldots,t_{i_{N_B}}\}$. Then, the forward model is applied on these frames only and the parameters are updated to minimize the difference between predicted data and the measured data $\{f_{t_{i_1}},\ldots, f_{t_{i_{N_B}}}\}$. This represents considerable benefits in terms of memory since the whole scene is never explicitly represented in the whole space-time grid and the cost of applying the forward operator on many frames is avoided, however, it adds variability during optimization.

Since neural fields are mesh-free, the regularization terms can be evaluated at any point of the domain. Additionally, derivatives can be computed through automatic differentiation. This motivates approximating the integrals for the regularizers in equation \eqref{eq:regularizers} via Monte Carlo by randomly sampling $N_C$ collocation points of the form $\{(x_c,t_c)\}_{c=1,\ldots,N_C}\subset\Omega_T$. For ease of notation, we introduce the function $\eta$ defined as the integrand in the regularization terms:
\[\begin{aligned}
\eta(u, \vv, x, t) := &\alpha\|\nabla u(x,t)\|_2+\beta\dsum_{j=1}^d\|\nabla v_j(x,t)\|_2\\
&+\gamma |\partial_tu(x,t)+\vv(x,t)\cdot\nabla u(x,t)|.
\end{aligned}\]
At a given iteration, frames and collocation points are randomly sampled and the parameters $\theta$ and $\phi$ of the networks are updated by minimizing the following function as an unbiased estimator of the objective \eqref{eq:variational problem}:
\[
\begin{array}{l}
\dfrac{1}{N_B}\dsum_{k=1}^{N_B}\mathcal{D}_{t_{i_k}}(u_{\theta}, f_{t_{i_k}}) + \dfrac{|\Omega_T|}{N_C}\dsum_{c=1}^{N_C}\eta(u_{\theta}, \vv_{\phi}, x_c, t_c).
\end{array}
\]
Finally, we recall that it is an open question how to choose $N_C$, the number of collocation points sampled at each iteration. One would like to sample as many points as possible to have a better approximation of the regularizer, however, this might be time-consuming and prohibitive in terms of memory because of the use of auto differentiation. Thus, we define the \emph{sampling rate} ($SR$) as the ratio between these collocation points and the amount of points on the spatiotemporal grid:
\begin{equation}\label{eq:sampling rate}
SR := \dfrac{N_C}{N_T\times N}.
\end{equation}



\subsection{Numerical evaluation with grid-based representation}\label{sec:grid-based method}

In this section we briefly describe the numerical realization of the grid-based representation of \eqref{eq:variational problem} as in \cite{Burger_Dirks_Schonlieb_2018}. A uniform grid $\{(x_j,t_i)\}_{i=1,\ldots,N_T; j=1,\ldots,N}\subset \Omega_T$ is assumed. Next, the quantities of interest are vectorized as $u\in\R^{N_T\times N}$, $\vv\in\R^{N_T\times N \times d}$, such that, $u_{ij}$ denotes the value of $u$ at the point $(x_j,t_i)$. The evaluation of the data fidelity term is now straightforward using \emph{Tomosipo}. For the regularization part finite difference schemes are employed to compute the corresponding derivatives. We let $D$ and $D_t$ denote the discretized gradients in space and time respectively (these could be forward or centred differences). Thus $(Du)\in \R^{N_T\times Nd}$, $(D_tu)\in\R^{N_T\times N}$, and $(Dv_j)\in \R^{N_T\times N\times d}$ for $j=1,\ldots, d$. Thus, the regularizers in \eqref{eq:regularizers} are approximated as follows:
\begin{equation*}
\begin{aligned}
\overline{\mathcal{R}}(u)&:=\dfrac{|\Omega_T|}{N_T N}\dsum_{i=1}^{N_T} \|(D u)_{i}\|_{2,1}, \\
\overline{\mathcal{A}}(r(u,\vv)) &:=\dfrac{|\Omega_T|}{N_T N}\dsum_{i=1}^{N_T} \|(D_t u)_{i} + \vv_{i}\cdot (D u)_{i}\|_1,\\
\overline{\mathcal{S}}(v)&:=\dfrac{|\Omega_T|}{N_T N} \dsum_{i=1}^{N_T} \dsum_{j=1}^d \| (D v_{j})_{i} \|_{2,1},
\end{aligned}
\end{equation*}
where $\|\cdot\|_{2,1}$ is a norm for vector fields, that first computes the 2-norm element-wise and then the 1-norm in space.

Using the previous, the variational problem \eqref{eq:variational problem} is discretized as 
\begin{equation}\label{eq:variational problem grid-based}
\begin{aligned}
\min_{u,\vv}&\mathcal{D}(u,\bold{f}) + \alpha\overline{\mathcal{R}}(u) + \beta \overline{\mathcal{S}}(\vv) + \gamma \overline{\mathcal{A}}(r(u,\vv)),
\end{aligned}
\end{equation}

This problem is non-convex due to the non-linearity present in the optical flow equation, however, it can be easily seen that it is biconvex, hence, in \cite{Burger_Dirks_Schonlieb_2018}, the proposed optimization routine updates the current iteration $(u^k,\vv^k)$ by alternating between the following two subproblems:

\begin{itemize}

\item Problem in $u$. Fix $\vv^k$ and update $u$ according to:
\begin{equation}\label{eq:problem in u}
\begin{aligned}
u^{k+1}=\argmin_{u} & \mathcal{D}(u,\bold{f}) + \alpha\overline{\mathcal{R}}(u) + \gamma \overline{\mathcal{A}}(r(u,\vv^k))
\end{aligned}
\end{equation}

\item Problem in $\vv$. Fix $u^{k+1}$ and update $\vv$ according to:
\begin{equation}\label{eq:problem in v}
\begin{aligned}
\vv^{k+1}=\argmin_{\vv}&\beta \overline{\mathcal{S}}(\vv) + \gamma \overline{\mathcal{A}}(r(u^{k+1},\vv))
\end{aligned}
\end{equation}
\end{itemize}

Each subproblem is convex with non-smooth terms involved that can be solved using the Primal-Dual Hybrid Gradient (PDHG) algorithm \cite{Chambolle_Pock_2010}. We refer to \cite{Burger_Dirks_Schonlieb_2018} for further details.

\section{Datasets}\label{sec:datasets}

We study our method in a 2D+time setting. We employ three synthetic datasets, the two-square, the cardiac, and the XCAT phantoms, and a real one, the \emph{STEMPO} phantom \cite{stempo}. For all the experiments the considered physical domain is the square $\Omega=[-1,1]^2$. Code will be made available on Github upon acceptance of the paper. 

For the two-square and cardiac phantoms, we define the ground-truth phantom $u:\Omega_T\to [0,1]$ as follows: 
\begin{itemize}
\item Let $u_0:\Omega\to[0,1]$ be the initial frame, i.e., we let $u(\cdot,0):=u_0(\cdot)$.
\item Define $\bold{\varphi}:\Omega_T\to\Omega$ describing the motion of the process. It takes a point $(x_0, y_0,t)\in\Omega_T$ and outputs $\varphi(x_0,y_0,t)\in\Omega$, the new position of $(x_0,y_0)$ at time $t$. For each time we can define the function $\bold{\varphi}_t:\Omega\to\Omega$ by $\bold{\varphi}_t(x_0,y_0)=\bold{\varphi}(x_0, y_0, t)$. We require $\bold{\varphi}_t$ to be a diffeomorphism for every $t\in[0,T]$. Hence we can define the trajectory of the point $(x_0,y_0)$ as $t \to \bold{\varphi}_t(x_0, y_0)$.
\item Define $u(x,y,t):= u_0(\bold{\varphi}_t^{-1}(x,y))$.
\end{itemize}
The phantom $u$ generated by the above procedure solves the optical flow equation with velocity field $\vv=\frac{d}{dt}\bold{\varphi}$. To mimic the continuous world, we define the ground truth at a high spatial resolution of $1024\times 1024$. Measurements are generated using \emph{Tomosipo}, assuming a camera with $M=64$ sensors. To avoid the inverse crime, during the reconstruction, we evaluate the neural field at a lower resolution spatial grid to get the discretized image as described in section \ref{sec:neural field method}. 

We shall consider two sampling strategies: random and sequential. For the first, at each frame a projection is taken along a random angle $\theta\in[0, 2\pi)$; for the second, projections are acquired at fixed 9-degree intervals between consecutive frames. For STEMPO, both strategies are sequential, the first one with 32-degree intervals and the second with 4-degree intervals between frames.

\subsection{Two-square phantom}

\begin{figure*}[]
\centering
\subfloat[First row: ground truth image at frames 15, 35, 55, 75, 95 (out of 100). Second row: velocity field at frames 15, 35, 55, 75, 95 (out of 100).]{\includegraphics[width=\textwidth]{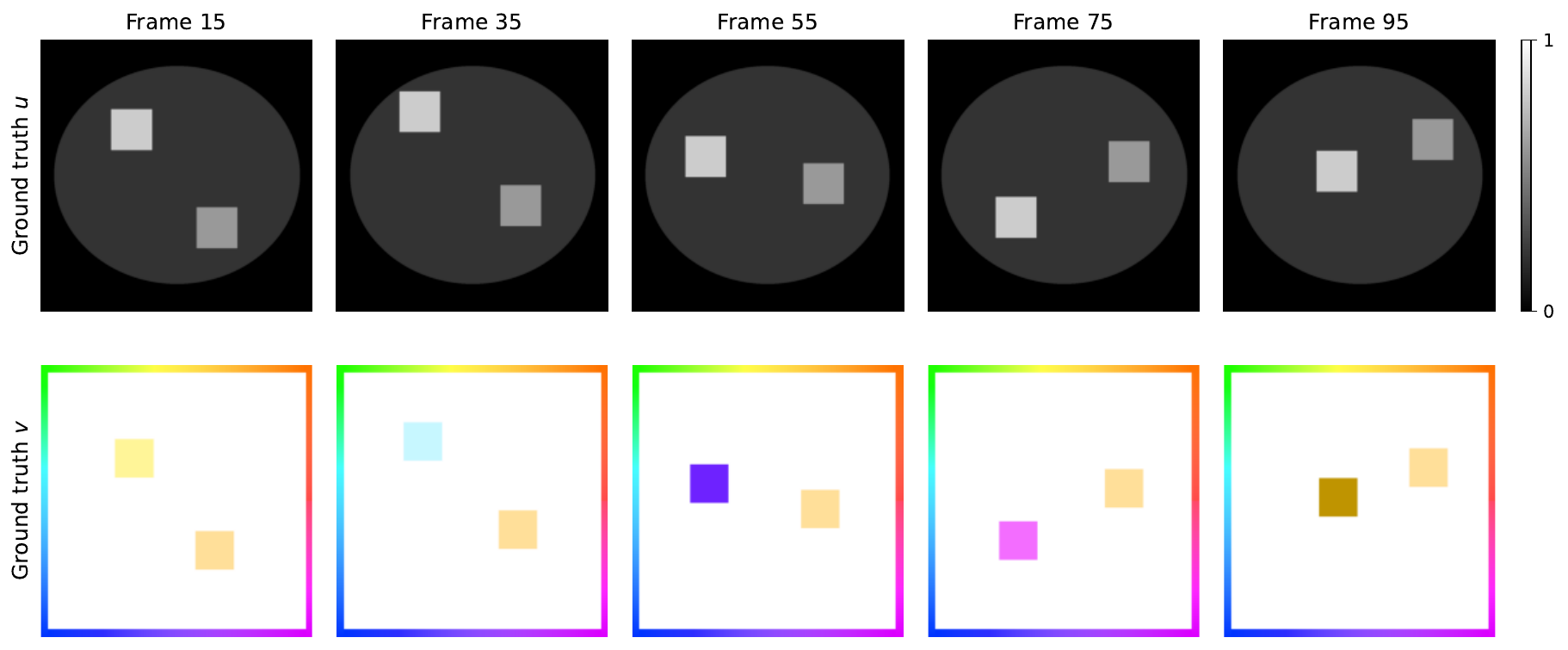}\label{fig:two-square ground truth}}
\vspace{0.5cm}
\subfloat[Fan-beam measurements $\mathbf{f}$. Left: random sampling. Right: sequential 9-degree sampling. $x$-axis denotes the time at which measurements were taken.]{\includegraphics[width=0.7\textwidth]{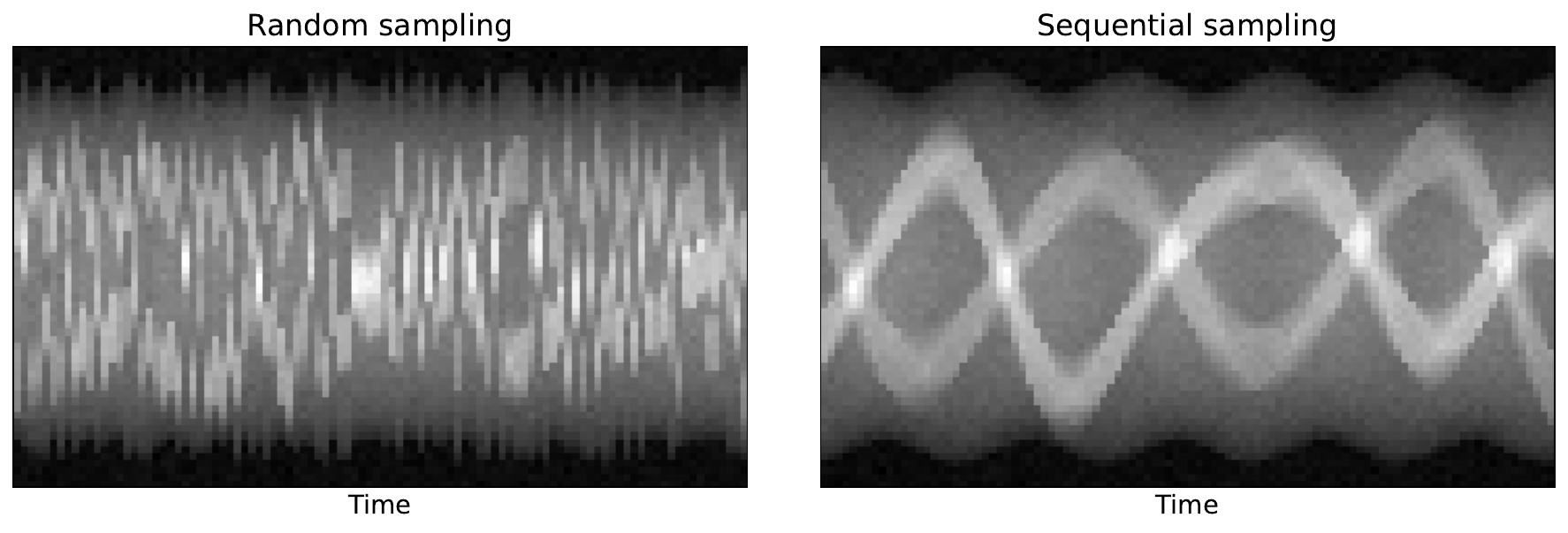}\label{fig:two-square measurements}}
\caption{Two-square phantom.}
\end{figure*}

The first phantom is depicted in figure \ref{fig:two-square ground truth} with two squares moving within an ellipsis-shaped background. The inverse of the motion for the squares on the left and right are $\varphi_1^{-1}$ and $\varphi_2^{-1}$ respectively, each one given by the following expressions:
\[\begin{aligned}
\varphi_1^{-1}(x,y,t) &= \begin{pmatrix}x-\frac{t}{5}\cos(2\pi t) \\ y-\frac{3t}{4}\sin(2\pi t) \end{pmatrix}, \\ 
\varphi_2^{-1}(x,y,t) &= \begin{pmatrix}x-0.3t \\ y-0.8t \end{pmatrix}.
\end{aligned}\]
From this, the velocity fields are easily expressed as:
\[\begin{aligned}
v_1(x,y,t)&=\begin{pmatrix}\frac{1}{5}\cos(2\pi t)-\frac{2\pi t}{5}\sin(2\pi t) \\ \frac{3}{4}\sin(2\pi t)+\frac{3\pi t}{2}\cos(2\pi t)\end{pmatrix}, \\
v_2(x,y,t)&=\begin{pmatrix}
0.3\\0.8
\end{pmatrix}.
\end{aligned}\]
$v_1$ produces a spiral-like motion for the square on the left and $v_2$ a constant diagonal motion for the square on the right. These are depicted in the second row of figure \ref{fig:two-square ground truth} as follows: the colored boundary frame indicates the direction of the velocity field. The intensities of the image indicate the magnitude of the vector. As an example, the square on the right moves constantly up and slightly to the right during the motion.

For this phantom, we set the distance source-origin to 3, the distance source-detector to 5, and the detector size to 3.5 to specify the projection geometry. We consider $N_T=100$ frames, leading to a measurement array of dimension $64\times100$. Measurements are further corrupted with Gaussian noise with standard deviation $0.01$. See figure \ref{fig:two-square measurements}. During reconstruction, each frame is generated by evaluating the neural field at a spatial grid of resolution $64\times 64$.

\begin{figure*}[]
\centering
\subfloat[First row: ground truth image at frames 1, 16, 25, 30, 70 (out of 300). Second row: velocity field at frames 1, 16, 25, 30, 70 (out of 300). Frames depict one cycle where everything shrinks and expands.]{\includegraphics[width=\textwidth]{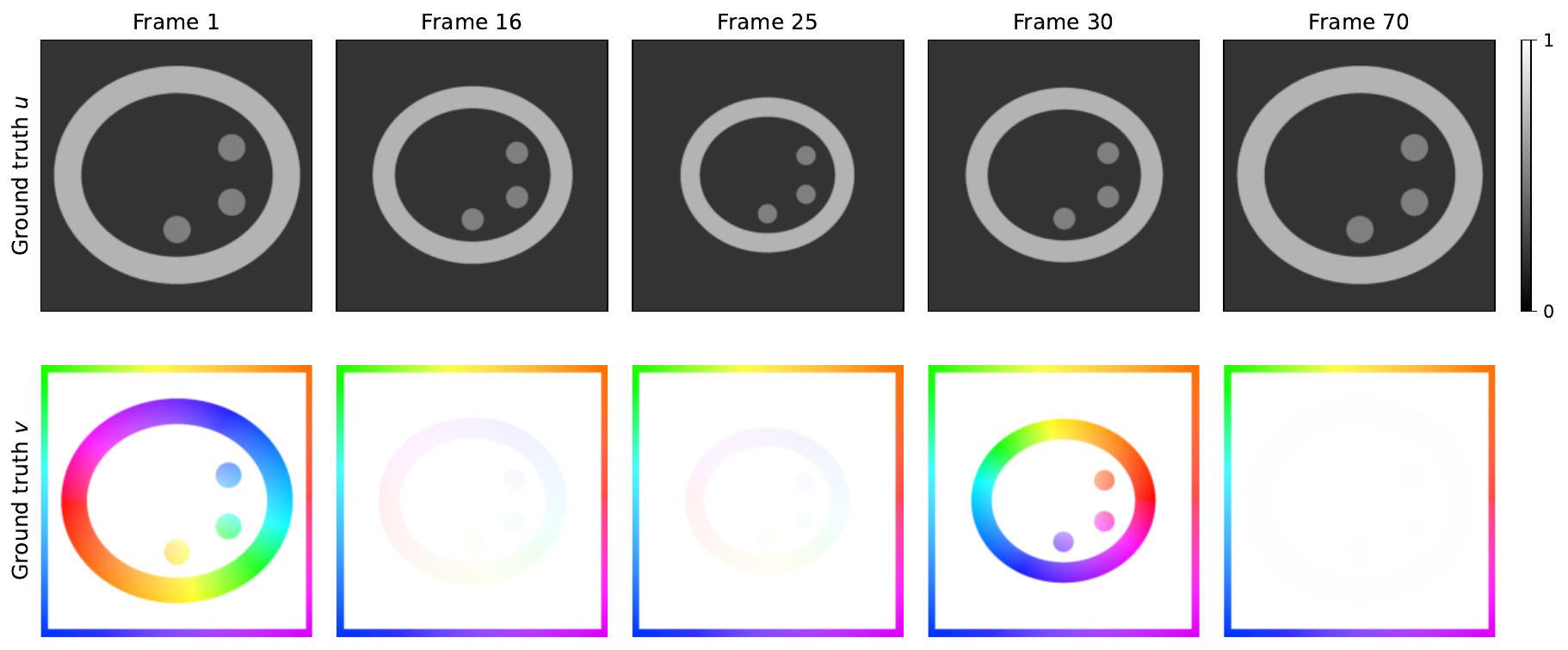}
\label{fig:cardiac ground truth}}
\vspace{0.5cm}
\subfloat[Fan-beam measurements $\mathbf{f}$. Left: random sampling. Right: sequential 9-degree sampling.]{\includegraphics[width=0.7\textwidth]{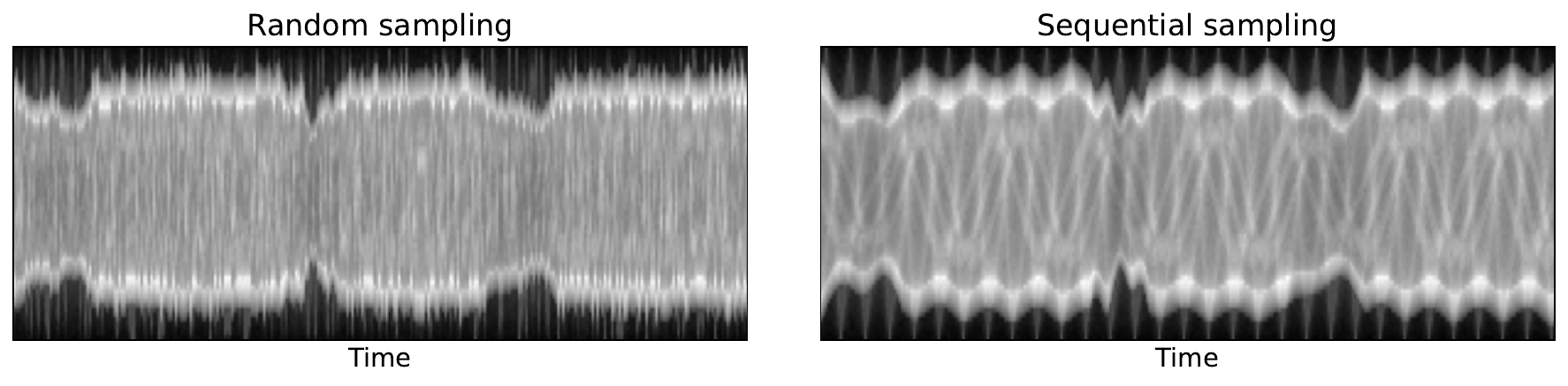}\label{fig:cardiac measurements}}
\vspace{0.5cm}
\subfloat[Function $a(t)$ describing the motion. The vertical lines at $t=1.1$ and $t=1.9$ indicate the beginning of a new period.]{\includegraphics[width=0.3\linewidth]{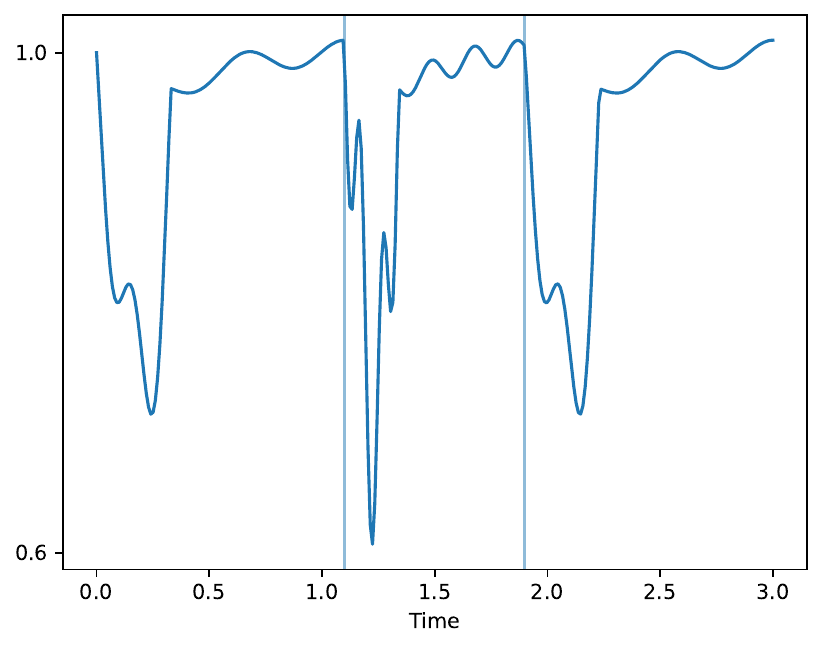}
\label{fig:alpha}}
\caption{Cardiac phantom.}
\end{figure*}

\subsection{Cardiac phantom}

Our second phantom shown in figure \ref{fig:cardiac ground truth} aims to mimic a heart-like motion. A slice of the heart is represented with an ellipse and three more circular-shaped structures are included. We also consider three cycles and set the final time to $T=3$. The motion is radial and described as follows:
\[\varphi(x_0,y_0,t) = a(t)\begin{pmatrix}x_0 \\y_0 \end{pmatrix},\]
where the function $t\to a(t)$ is depicted in figure \ref{fig:alpha}. As can be seen, it consists of three periods, with the first and third periods following the same pattern, while the second one depicts an intricate and irregular motion resembling an arrhythmia.

We set the same projection geometry as for the two-square phantom. For this phantom we consider $N_T=300$ frames, leading to a measurement array of dimension $64\times300$. Measurements are further corrupted with Gaussian noise with standard deviation $0.01$. See figure \ref{fig:cardiac measurements}. During reconstruction, each frame is generated by evaluating the neural field at a spatial grid of resolution $64\times 64$.

\subsection{STEMPO}

In \cite{stempo}, the \emph{Spatio-TEmporal Motor-Powered} (STEMPO) dataset is introduced. It provides X-ray tomography data for a moving object. The dynamics of the object are controlled by a motor, allowing for different sampling schemes to be performed. The phantom consists of two static objects and a square moving upward from the bottom to the top at a constant speed. 

The whole phantom is fully sampled at its initial and static state. A ground truth for the initial frame is then obtained by taking a filtered back projection and then setting to 0 those pixels below a given threshold. This is done to reduce the effect of the noise in the measurements and provide a clean background for the image. During the motion only one projection is acquired at a single time instance, collecting a total of 360 frames. Additionally, two sequential sampling schemes are provided, the first one takes projections every one degree, and the second one takes projections every 8 degrees. To generate the subsequent frames for the ground truth, the reconstructed initial frame is extrapolated according to the known motion given by the motor. 

The acquired data is then binned at different factors. Here we consider a binning factor of 32, leading to each projection having 70 measurements. Given the simplicity of the motion, we add another layer of difficulty by uniformly downsampling the number of measured frames from 360 to 90. This leads to projections every 4 degrees and 32 degrees. For this phantom, we also consider a domain $[-1,1]^2\times[0,1]$, for which the real geometry is accordingly scaled, for instance, the distance source-origin is 9.88, the distance source-detector is 13.33, and the size of the detector is 2.69. During reconstruction, each frame is generated by evaluating the neural field at a spatial grid of resolution $70\times 70$.
\begin{figure*}[]
\centering

\subfloat[STEMPO ground truth image at frames 1, 20, 40, 60, 90 (out of 90). The square moves from the bottom to the top. All other structures are static.]{\includegraphics[width=\textwidth]{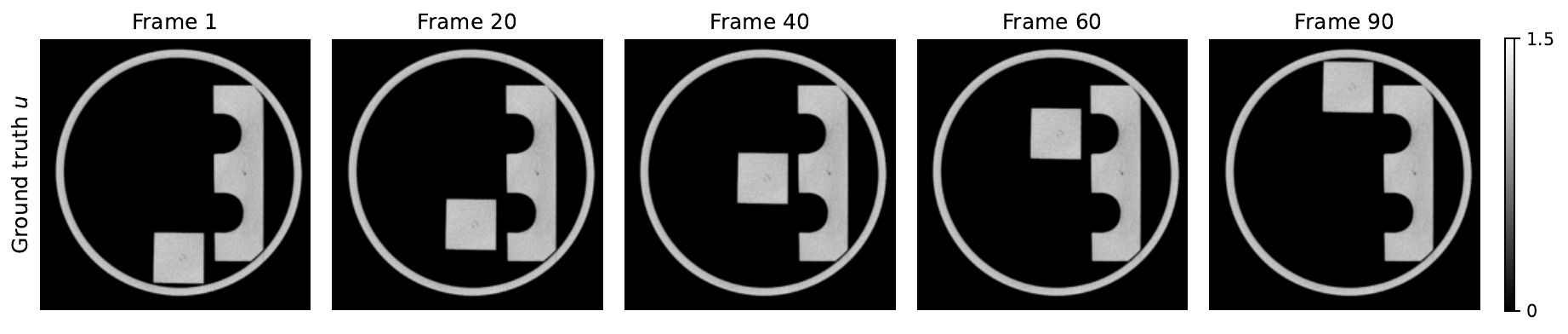}
\label{fig:stempo ground truth}}
\vspace{0.5cm}
\subfloat[Fan-beam measurements $\mathbf{f}$. Left: sequential 32-degree sampling. Right: sequential 4-degree sampling.]{\includegraphics[width=0.7\textwidth]{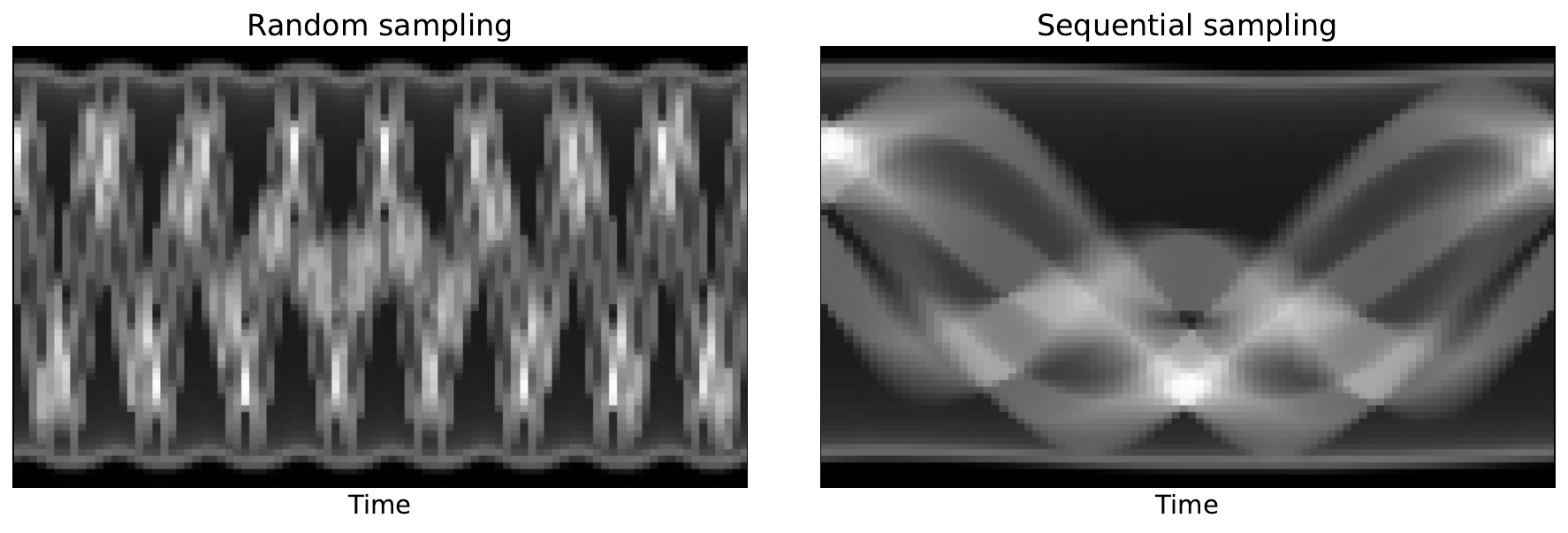}
\label{fig:stempo measurements}}
\caption{STEMPO dataset.}
\end{figure*}

\subsection{XCAT phantom}

A simulated thoracic phantom generated by the 4DXCAT software \cite{segars20104d} is used to assess our method on more complex structures with different motion magnitudes. This phantom has been made publicly available in \cite{huang2024resolving}\footnote{\url{https://rdr.ucl.ac.uk/articles/dataset/4DCT_XCAT_phantom_dataset_for_Resolving_Variable_Respiratory_Motion_From_Unsorted_4D_Computed_Tomography_MICCAI2024/26132077}}. The phantom is a 3D+time volume with 182 frames and $355\times 280 \times 115$ spatial voxels spanning 18 respiratory cycles. The phantom is zero padded to get $355\times 355 \times 115$ voxels.

\begin{figure*}[]
    \centering    \includegraphics[width=1\linewidth]{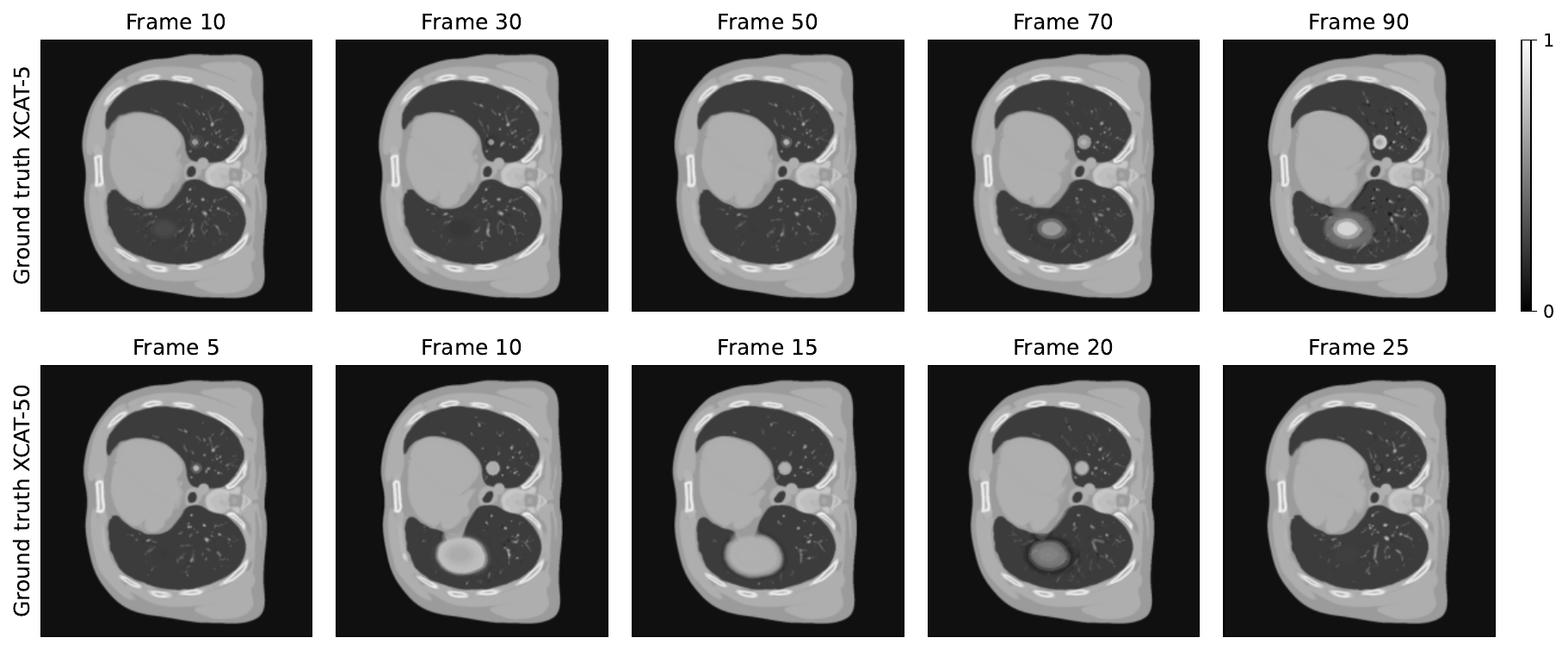}
    \caption{Top: XCAT-5 ground truth image at frames 10, 30, 50, 70, 90 (out of 100). XCAT-5 represents almost one respiratory cycle. Bottom: XCAT-50 ground truth image at frames 5, 10, 15, 20, 25 (out of 100). XCAT-50 represents almost five respiratory cycles. }
    \label{fig:xcat}
\end{figure*}

A natural question is the effect of the magnitude of the motion on the reconstruction quality. To assess this, the XCAT phantom is employed to generate 5 phantoms with different motions as follows: i) we select the first 5, 10, 20, 35, and 50 frames from the original XCAT phantom, ii) use cubic interpolation in time to get $N_T=100$ frames for each phantom, and iii) select the slice $z=40$ to get a 2D+time reconstruction problem. We refer to these phantoms as XCAT-$j$, for $j=5,10,20,35,50$. Each respiratory cycle lasts approximately 10 frames of the original phantom, thus XCAT-5 represents half of a cycle, while XCAT-50 represents 5 cycles. Intuitively, the reconstructed image is expected to worsen as the motion increases. An additional challenge of this phantom is that it presents out-of-plane motion with the diaphragm coming in and out of the slice, meaning that the optical flow equation is not satisfied there. Figure \ref{fig:xcat} depicts one respiratory cycle of the XCAT-5 and XCAT-50 phantoms.

We now proceed to specify the projection geometry for these phantoms. We set the distance source-origin to 6, the distance source-detector to 8, and the detector size to 3.5. We consider $N_T=100$ frames, and a camera with $M=150$ sensors, leading to a measurement array of dimension $150\times100$. Measurements are further corrupted with Gaussian noise with standard deviation $0.005$. During reconstruction, each frame is generated by evaluating the neural field at a spatial grid of resolution $150\times 150$.

\section{Numerical Experiments}\label{sec:numerical experiments}

As mentioned before, in our numerical experiments, we use \emph{Tomosipo} to compute the X-ray transform and its transpose. For both $u_{\theta}$ and $\vv_{\phi}$ we use $m_x=m_t=32$ for the Fourier mappings \eqref{eq:architecture}. Hence, both neural fields embed $(x,t)$ into a vector of size $m=128$. We then set $L=3$ hidden layers with $d_l=128$ neurons each. We use $\tanh$ as activation function. Notice that we do not apply an activation function in the last layer, in particular, we do not impose the network $u_{\theta}$ to output non-negative values. 

We find $\sigma_x=\sigma_t=0.1$ to give the best results for the two-square and the STEMPO datasets, $\sigma_x=0.1, \sigma_t=0.5$ for the cardiac dataset, and $\sigma_x=\sigma_t=0.5$ for the XCAT-$j$ phantoms. We try two batch settings, the full-batch $N_B=N_T$ and the mini-batch $N_B=1$. In the full-batch setting we optimize for $150,000$ iterations, while in the mini-batch setting, we optimize for $10,000$ epochs (leading to $10,000\times N_T$ iterations). For the collocation points, we use a Latin Hypercube Sampling strategy \cite{stein1987large}. We set the sampling rate \eqref{eq:sampling rate} at $SR=0.1$, for the two-square, cardiac, and STEMPO datasets, and $SR=0.01$ for the XCAT-$j$ datasets, leading to $40,960$, $122,880$, $44,100$, and $22,500$ collocation points being randomly sampled in each iteration, respectively. We use the Adam optimizer with a learning rate of $10^{-3}$ in all cases. For the grid-based approach, each subproblem, \eqref{eq:problem in u} and \eqref{eq:problem in v}, are run for $2,000$ iterations and this alternation is repeated 5 times. The PSNR values reported in this section are computed using the resolution during reconstruction. Since the ground truth is defined at a higher resolution in space, we downsample it using a pooling average. Experiments are performed on an Nvidia Tesla V100 GPU with 16GB \cite{https://doi.org/10.15125/b6cd-s854}. In this paper, we are not concerned with the computational speed of the methods. Note that our naive implementation takes around a few hours for each data set (see figure \ref{fig:unregularized versus regularized}).



\subsection{Regularization of  neural fields}\label{sec:unreg vs reg}

\begin{figure*}[]
\centering
\subfloat[Two-square phantom]{\includegraphics[width=0.32\textwidth]{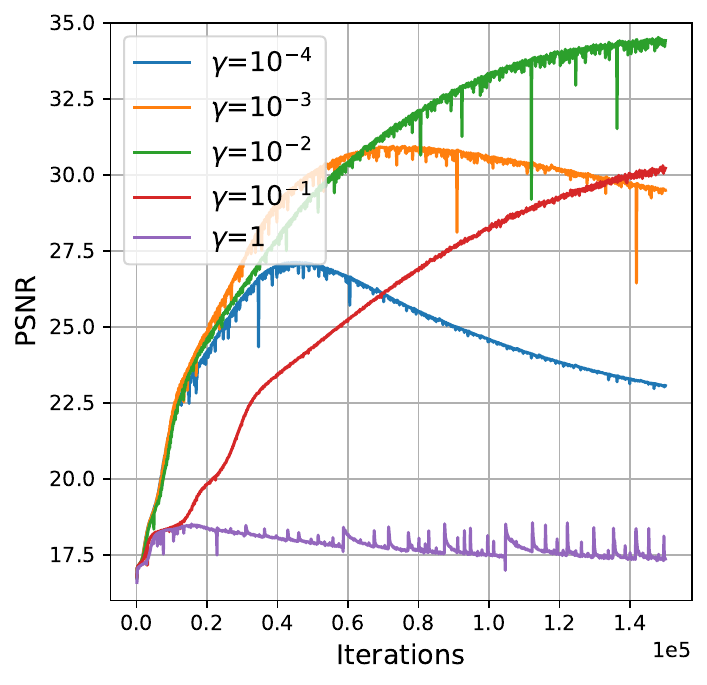}\label{fig:two-square varying gamma}}
\subfloat[Cardiac phantom]{\includegraphics[width=0.32\textwidth]{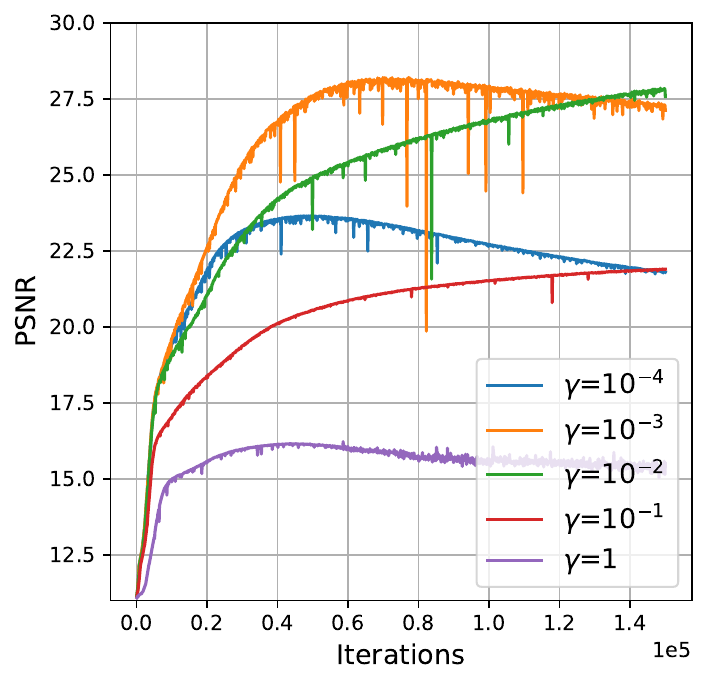}\label{fig:cardiac varying gamma v2}}
\subfloat[STEMPO phantom]{\includegraphics[width=0.32\textwidth]{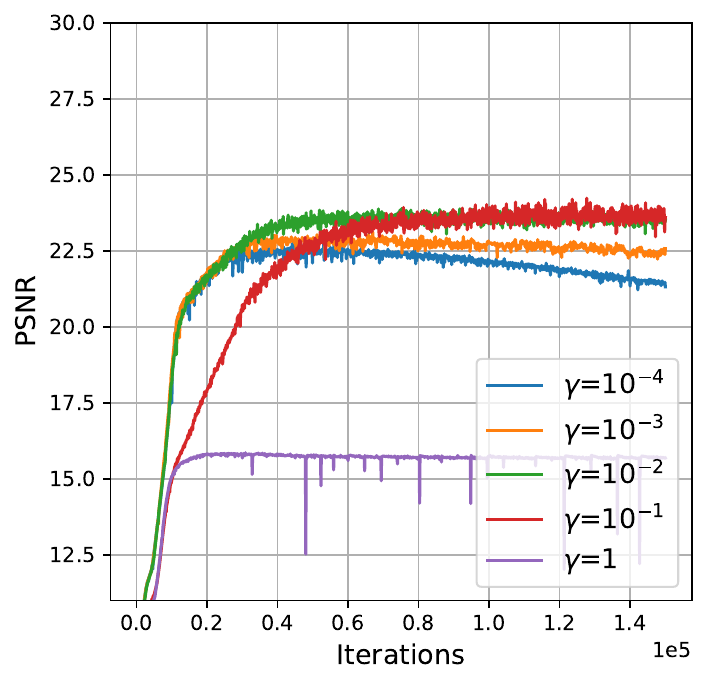}\label{fig:stempo varying gamma}}
\caption{Ablation study on $\gamma$ for neural fields. Each plot shows the evolution of PSNR during optimization for $\gamma\in\{10^{-4}, 10^{-3}, 10^{-2}, 10^{-1}, 1\}$ and with batch size $N_B=N_T$. $\gamma=10^{-2}$ achieves superior performance in the three phantoms.}
\label{fig:varying gamma}

\vspace{0.2cm}
\subfloat[Two-square phantom]{\includegraphics[width=0.32\textwidth]{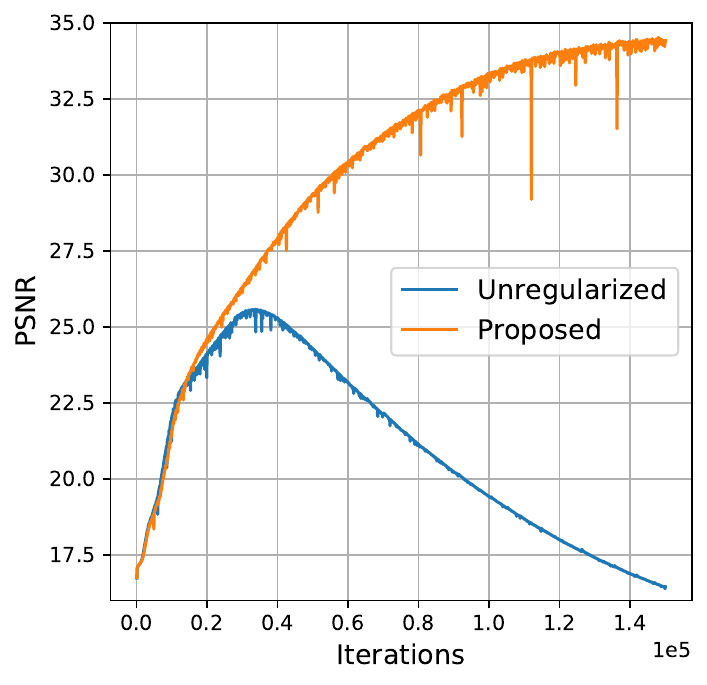}\label{fig:two-square unregularized versus regularized}}
\subfloat[Cardiac phantom]{\includegraphics[width=0.32\textwidth]{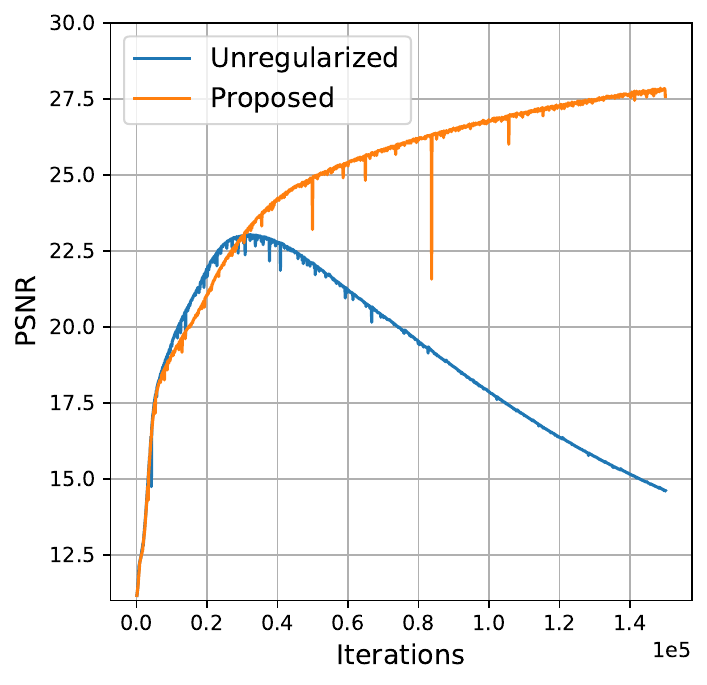}\label{fig:cardiac unregularized versus regularized}}
\subfloat[STEMPO phantom]{\includegraphics[width=0.32\textwidth]{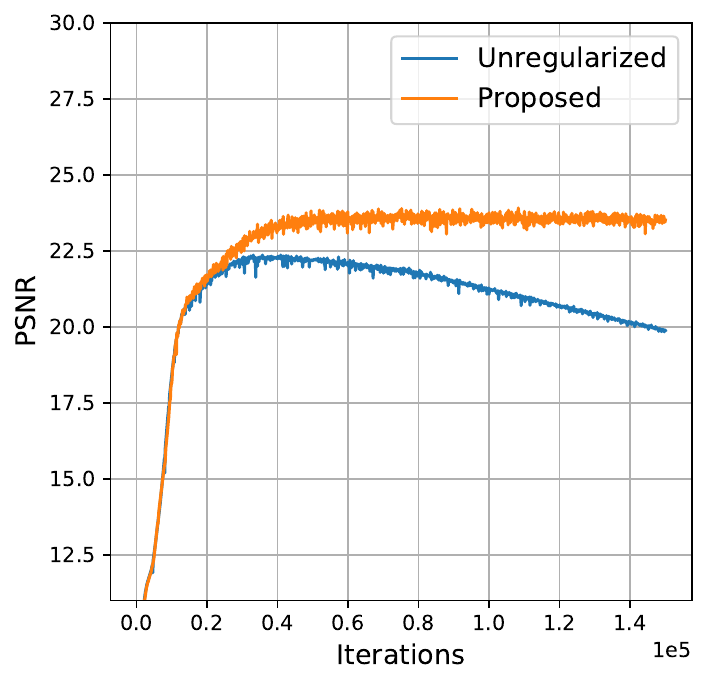}\label{fig:stempo unregularized versus regularized}}
\caption{Evolution of PSNR during optimization for the unregularized ($\gamma=0$) and the proposed regularized solution $(\gamma=10^{-2})$ with batch size $N_B=N_T$. The proposed regularized solution achieves a higher PSNR and avoids fitting noise.}
\label{fig:unregularized versus regularized}

\vspace{0.2cm}
\subfloat[Two-square phantom]{\includegraphics[width=0.32\textwidth]{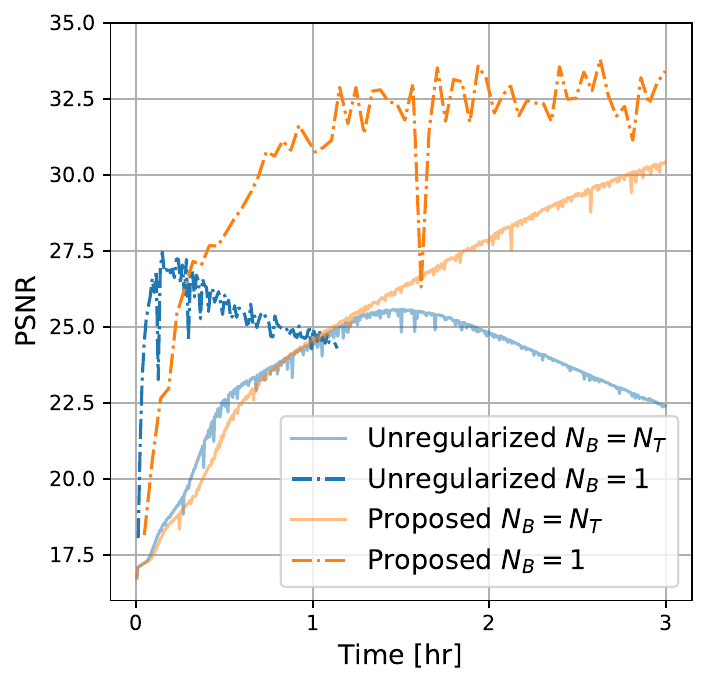}\label{fig:two-square batch size}}
\subfloat[Cardiac phantom]{\includegraphics[width=0.32\textwidth]{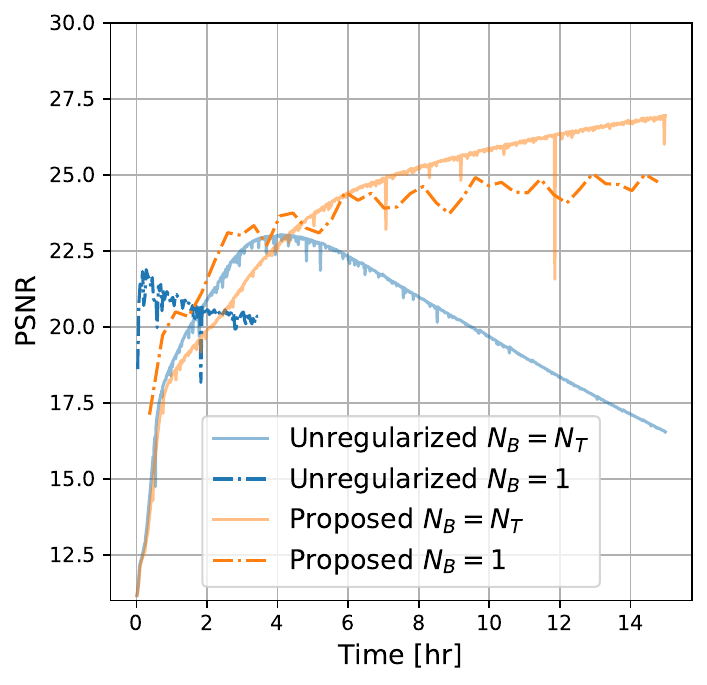}\label{fig:cardiac batch size}}
\subfloat[STEMPO phantom]{\includegraphics[width=0.32\textwidth]{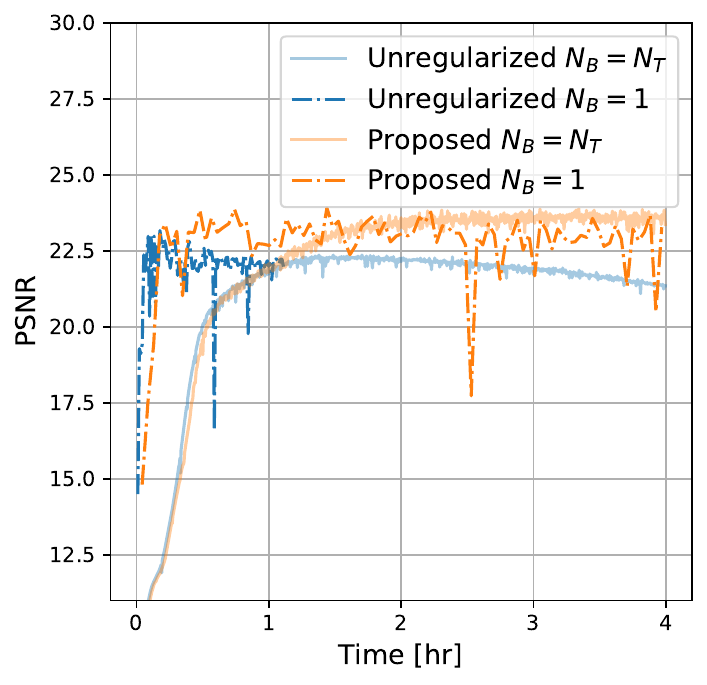}\label{fig:stempo batch size}}
\caption{Comparing training in terms of batch size $N_B\in\{1, N_T\}$ and $\gamma\in\{0, 10^{-2}\}$. Each plot shows the evolution of PSNR during optimization against time in hours. $N_B=1$ achieves a higher PSNR in less time but with high variability.}
\label{fig:batch size}
\end{figure*}

\begin{figure*}[b]
\centering
\includegraphics[width=\textwidth]{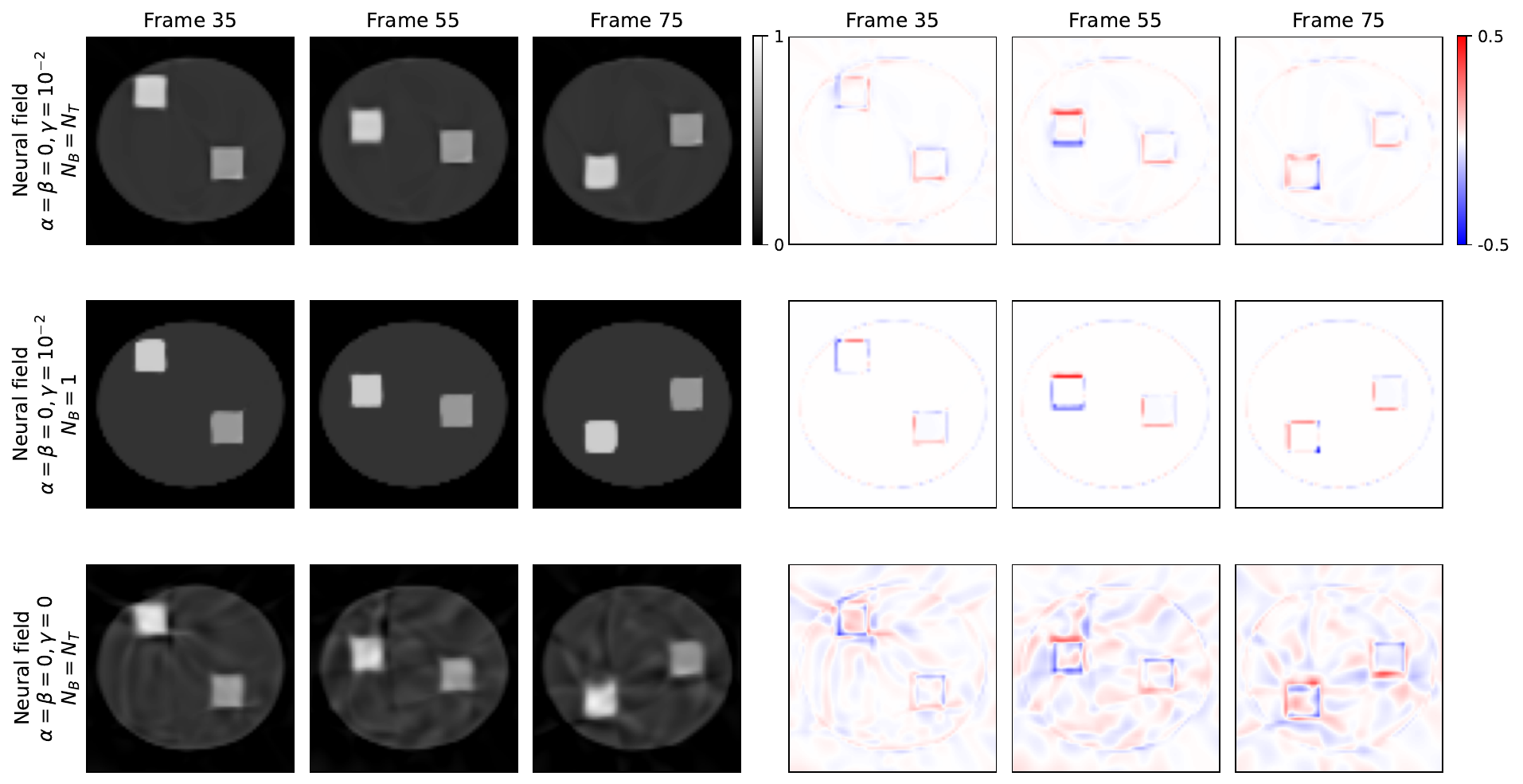}
\caption{Reconstruction (left) with neural fields and error (right) of the two-square phantom for the random sampling strategy at frames 35, 55, and 75 (out of 100). $\gamma=10^{-2}$ achieves smaller errors and $N_B = 1$ gets sharper edges.}
\label{fig:two-square reconstructions}
\end{figure*}

In this section, we study the effect of optimizing the neural field with an explicit motion regularizer and compare it against the purely implicitly regularized solution. For this purpose, we consider a simplification of the variational problem \eqref{eq:variational problem} by setting the regularization parameters $\alpha,\beta=0$ and performing an ablation study on $\gamma \in \{10^{-4},10^{-3},10^{-2},10^{-1}, 1\}$ in the two-square, cardiac, and STEMPO datasets with random sampling. This choice helps to understand the role of the optical flow by isolating it from the other regularizers $\mathcal{R}$ and $\mathcal{S}$.


In figure \ref{fig:varying gamma} we show the evolution of PSNR during optimization for the three datasets. This figure highlights the role played by $\gamma$ in the reconstruction. On the one hand, for the lowest value, $\gamma=10^{-4}$, it can be seen a semi-convergence behavior where approximately after 40,000 epochs the neural field starts fitting the noise in the measurements and the reconstruction quality decreases steadily throughout the optimization. On the other hand, for the highest value, $\gamma=1$, the regularization term is too strongly imposed and a completely static image is obtained. For the two-square phantom, $\gamma=10^{-2}$ gives the best result in terms of PSNR achieving a value of 34.41 after 150,000 iterations; for the cardiac phantom, $\gamma=10^{-3}$ performs the best during most of the optimization, however, its quality starts decreasing after 60,000 iterations and the case $\gamma=10^{-2}$ achieves a higher PSNR of 28.03 at the last iteration; for STEMPO, we can see that both $\gamma=10^{-2}$ and $\gamma=10^{-1}$ achieve a similar PSNR but the former evolves more quickly and attains a value of 23.51. We therefore propose a regularized solution using $\gamma=10^{-2}$ for the three datasets.

We continue our study by comparing the unregularized case $\gamma=0$ against the proposed regularized one. This is shown in figure \ref{fig:unregularized versus regularized}. In all cases, $\gamma=0$ shows degradation on the reconstruction early during optimization, after around 30,000 iterations. Its regularized counterpart $\gamma=10^{-2}$ is superior across almost all the iterations. More importantly, this regularizer stabilizes the reconstruction and prevents the neural field from fitting the noise in the measurements.

We also compare the full-batch setting $N_B=N_T$ against the commonly used mini-batch setting with $N_B = 1$, and for $\gamma\in\{0, 10^{-2}\}$. We recall that for $N_B=1$ optimization is done for 10,000 epochs. Results for the evolution of PSNR versus time are shown in figure \ref{fig:batch size}. A batch size of 1 speeds up the optimization, allowing to achieve a higher PSNR in less time. It suffers however from an unstable behaviour with the PSNR varying for almost 2 points every 100 epochs. This means we can stop at a poor reconstruction without a reliable stopping criteria.

\begin{figure*}[]
\centering
\includegraphics[width=\textwidth]{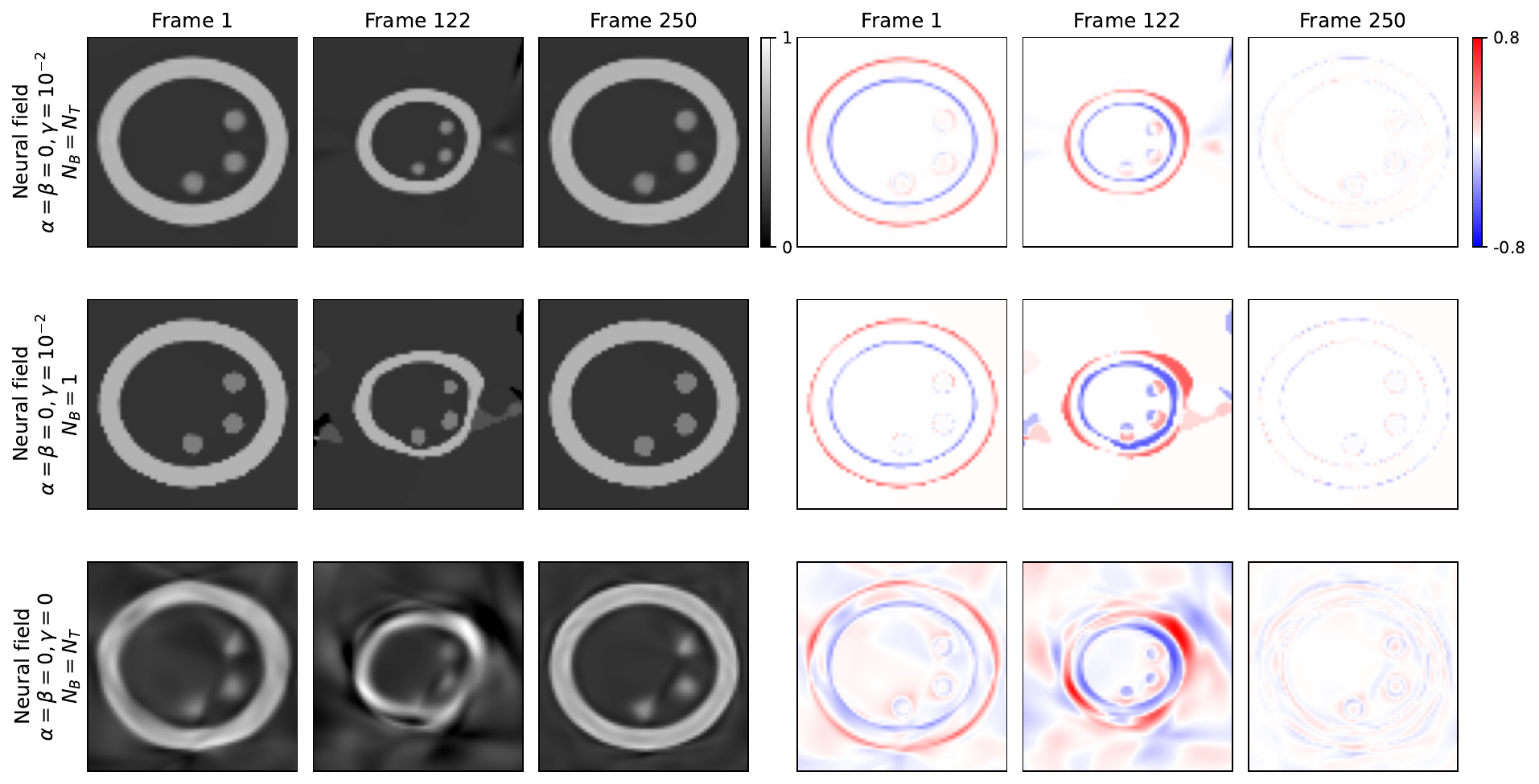}
\caption{Reconstruction (left) with neural fields and error (right) of the cardiac phantom for the random sampling strategy at frames 1, 122, and 250 (out of 300). $\gamma=10^{-2}$ achieves smaller errors and $N_B = 1$ gets sharper edges.}
\label{fig:cardiac reconstructions}
\vspace{0.5cm}
\centering
\includegraphics[width=\textwidth]{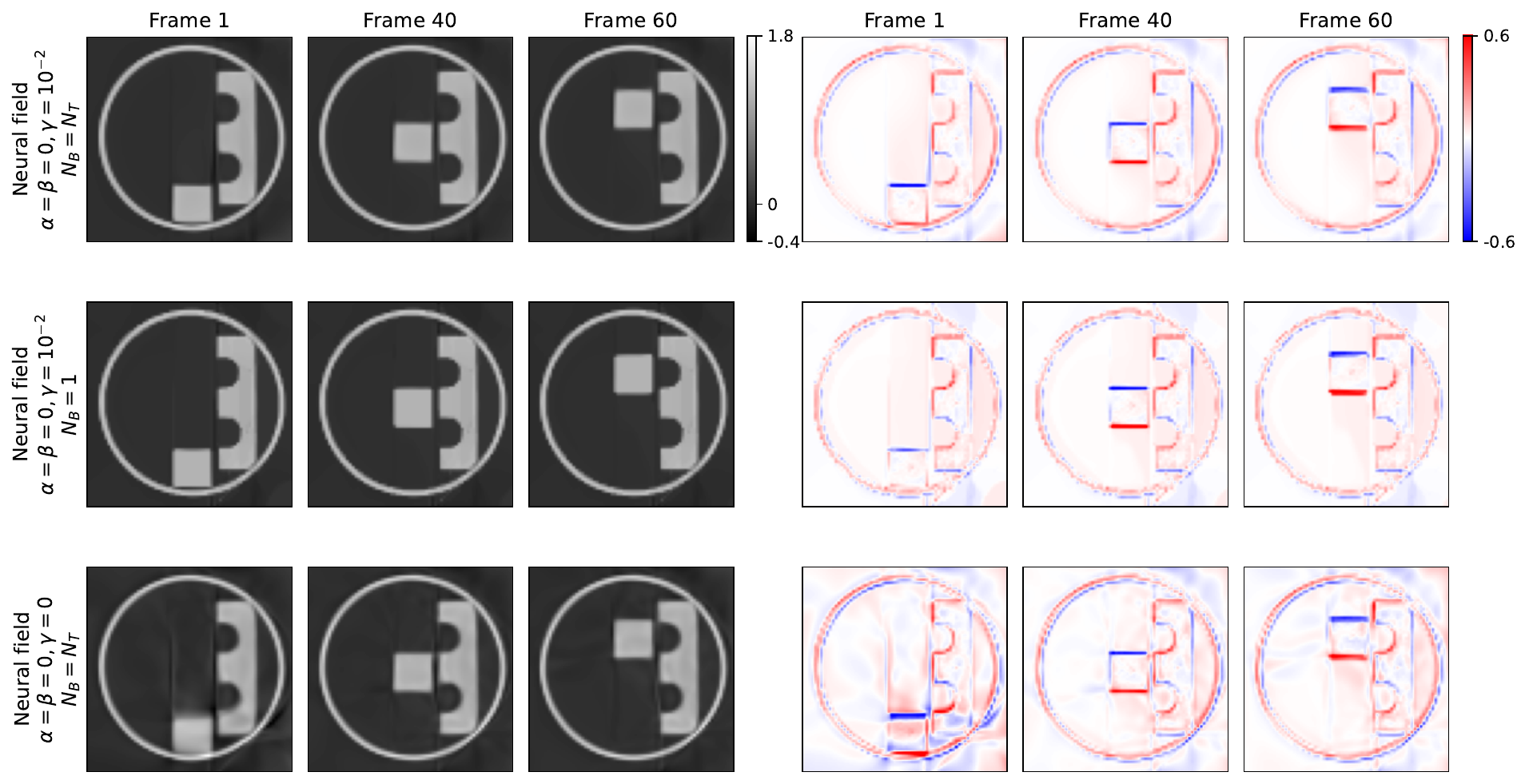}
\caption{Reconstruction (left) with neural fields and error (right) of the STEMPO phantom for the sequential 32-degree sampling strategy at frames 1, 40, and 60 (out of 90). $\gamma=0$ gets larger errors.}
\label{fig:stempo reconstructions}
\end{figure*}

Obtained reconstructions at their best PSNR and the corresponding error against the ground truth at different frames for the two-square, cardiac, and STEMPO datasets are displayed in figures \ref{fig:two-square reconstructions}, \ref{fig:cardiac reconstructions}, and \ref{fig:stempo reconstructions}, respectively. The top rows show the full-batch regularized case $N_B=N_T, \gamma=10^{-2}$; the middle rows show the mini-batch regularized case $N_B=1, \gamma=10^{-2}$; the bottom rows show the full-batch unregularized case $N_B=N_T, \gamma=0$.  A relevant advantage in the mini-batch setting is that it is more likely to capture edges in less time since more iterations can be taken, while its full-batch counterpart $N_B=N_T$ shows blurry edges. We also summarize the PSNR values obtained for the different models in table \ref{table:reg vs unreg}, highlighting the role of the PDE-based regularizer.

\begin{table}[h]
\centering
\begin{tabular}{@{}cc|ccc@{}}
\toprule
\multicolumn{2}{c|}{Parameters} & \multicolumn{3}{c}{Datasets} \\ \midrule
$\gamma$ & $N_B$ & Two-square & Cardiac & STEMPO \\ \midrule
\multicolumn{1}{|c|}{$10^{-2}$} & $N_T$ & \multicolumn{1}{c|}{\textbf{34.52}} & \multicolumn{1}{c|}{\textbf{28.09}} & \multicolumn{1}{c|}{\textbf{23.91}} \\ \midrule
\multicolumn{1}{|c|}{$10^{-2}$} & 1 & \multicolumn{1}{c|}{34.28} & \multicolumn{1}{c|}{25.32} & \multicolumn{1}{c|}{\textbf{23.91}} \\ \midrule
\multicolumn{1}{|c|}{0} & $N_T$ & \multicolumn{1}{c|}{25.58} & \multicolumn{1}{c|}{23.03} & \multicolumn{1}{c|}{22.36} \\ \midrule
\multicolumn{1}{|c|}{0} & 1 & \multicolumn{1}{c|}{27.46} & \multicolumn{1}{c|}{21.88} & \multicolumn{1}{c|}{23.16} \\ \bottomrule
\end{tabular}
\caption{Maximum PSNR attained during optimization obtained by neural fields with $\alpha=\beta=0$ and varying the motion regularization parameter $\gamma$ and the batch size $N_B$.}
\label{table:reg vs unreg}
\end{table}

\begin{figure*}[]
\centering
\subfloat[Effect of optical flow regularizer.]{\includegraphics[width=0.32\textwidth]{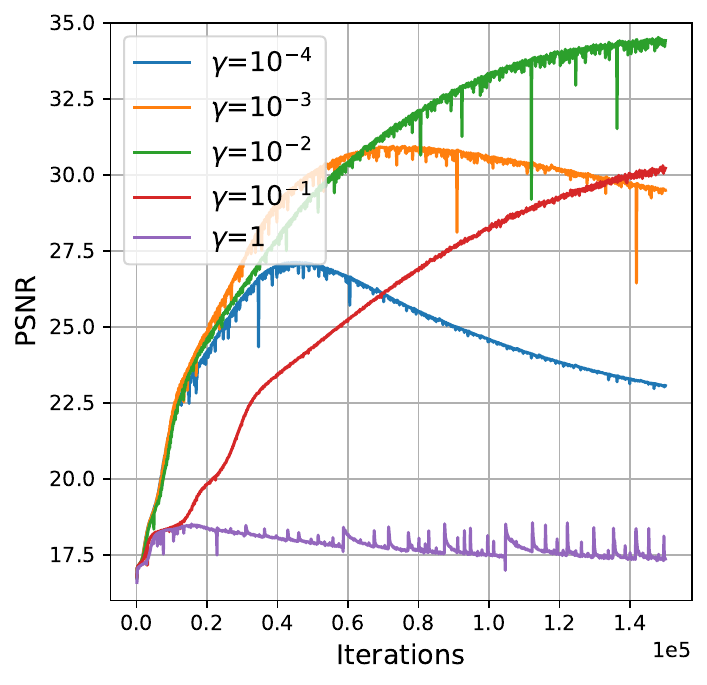}\label{fig:app optical flow}}
\subfloat[Effect of TV regularizer.]{\includegraphics[width=0.32\textwidth]{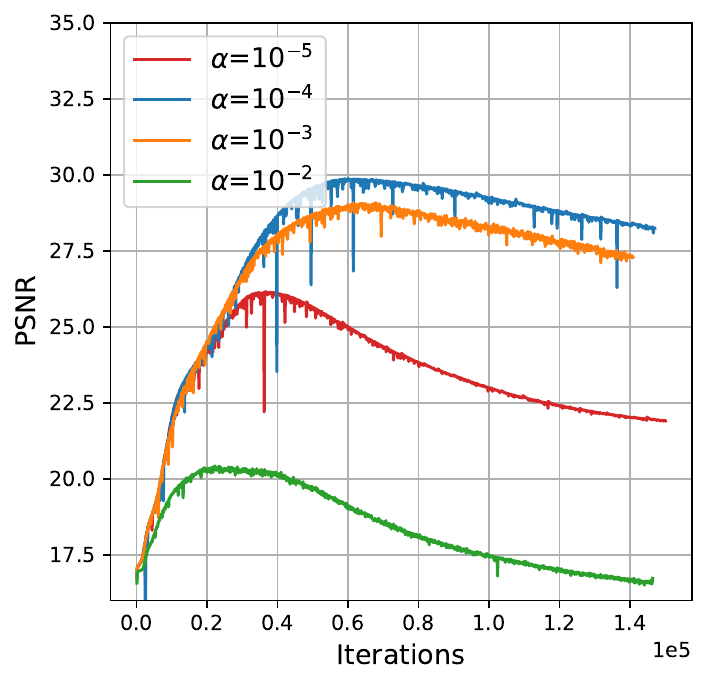}\label{fig:app TV}}
\subfloat[Effect of STV regularizer.]{\includegraphics[width=0.32\textwidth]{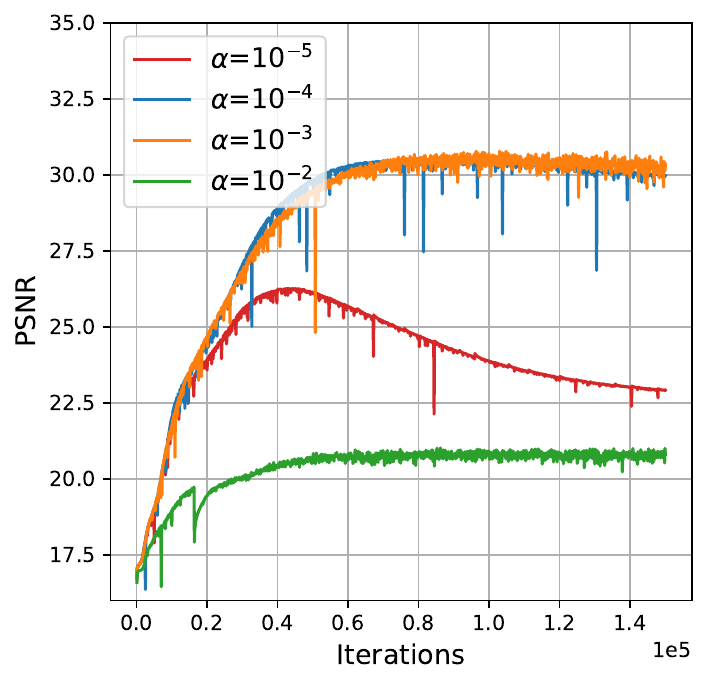}\label{fig:app STV}}
\caption{Comparing optical flow and TV-like regularizers on the two-square phantom. Left: ablation study on $\gamma$ while setting $\alpha,\beta=0$. Center: ablation study on $\alpha$ while setting $\beta,\gamma=0$ using the TV regularizer in space. Right: ablation study on $\alpha$ while setting $\beta,\gamma=0$ using the STV regularizer. A higher PSNR is attained for the optical flow regularizer. }
\label{fig:app regularizers}
\end{figure*}

We finish this section by comparing the PDE-based motion regularizer against the more common total variation and show the benefits of employing the former. To showcase the effects of the TV regularizer, we proceed in a similar way as in the previous experiment, this is, setting $\beta,\gamma=0$ and performing an ablation study on $\alpha$. Additionally, to complement our study, we consider a third regularizer, the spatiotemporal total variation (STV) regularizer, which differs from $\mathcal{R}$ in that it penalizes changes in time as well and is defined as follows:
\[\hat{\mathcal{R}}(u):=\dint_{\Omega_T}\sqrt{ \|\nabla u\|^2+ (\partial_t u)^2}.\]
We consider the two-square phantom for this experiment. Results displayed in figure \ref{fig:app regularizers} show that the best reconstruction using the optical flow constraint enhances the reconstruction task by almost 5 when compared to the best reconstruction using TV or STV regularization.

\subsection{Effects of rapid motion}

In this section we employ the XCAT-$j$ phantoms to study the effects of motion in the reconstruction. We recall that, by construction, the larger is $j$, the larger is the motion. For this experiment we set the Fourier feature hyperparameters $\sigma_x=\sigma_t=0.5$, the regularization parameters $(\alpha,\beta,\gamma) = (10^{-5}, 10^{-5}, 10^{-3})$, the batch size $N_B=1$, and train for 10,000 epochs for the five XCAT-$j$ phantoms. 

\begin{figure*}[]
    \centering
    \includegraphics[width=1\linewidth]{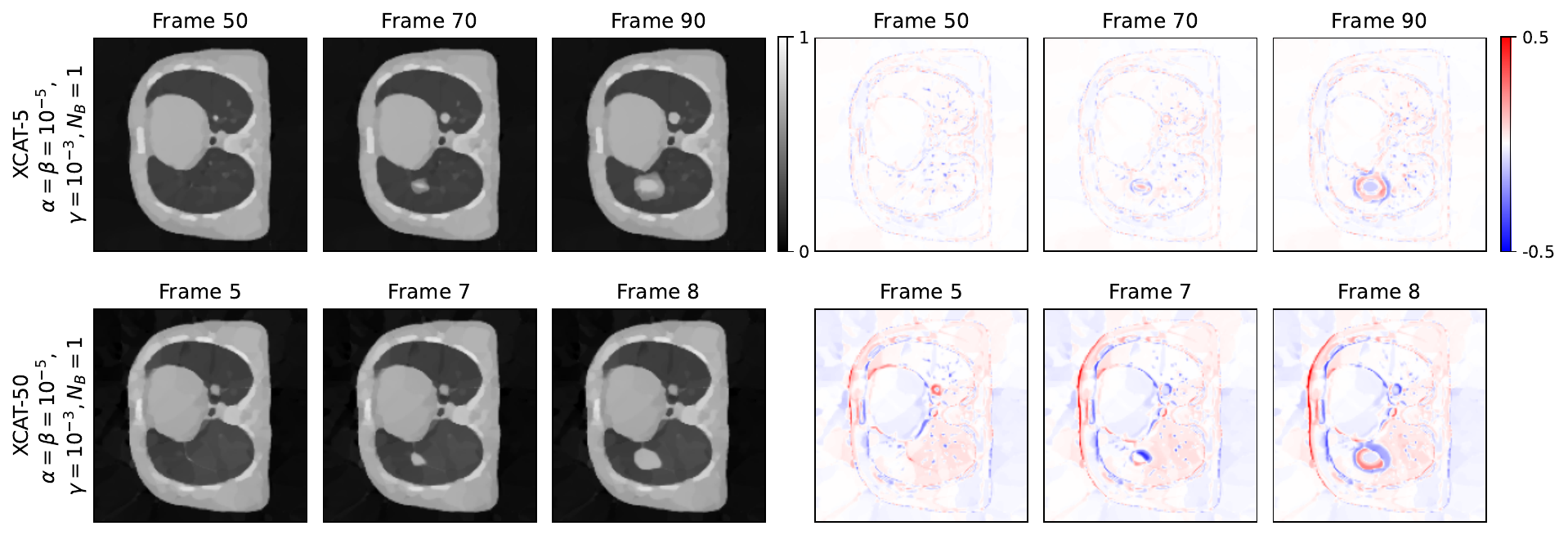}
    \caption{Reconstruction (left) with neural fields and error (right) for XCAT-5 at frames 50, 70, 90 and XCAT-50 at frames 5, 7, 8. Depicted frames for XCAT-5 and XCAT-50 represent similar time instances from the first respiratory cycle. Motion magnitude is larger in XCAT-50, thus, larger errors are obtained.}
    \label{fig:rec xcat}
\end{figure*}

\begin{figure}[]
    \centering
    \includegraphics[width=1\linewidth]{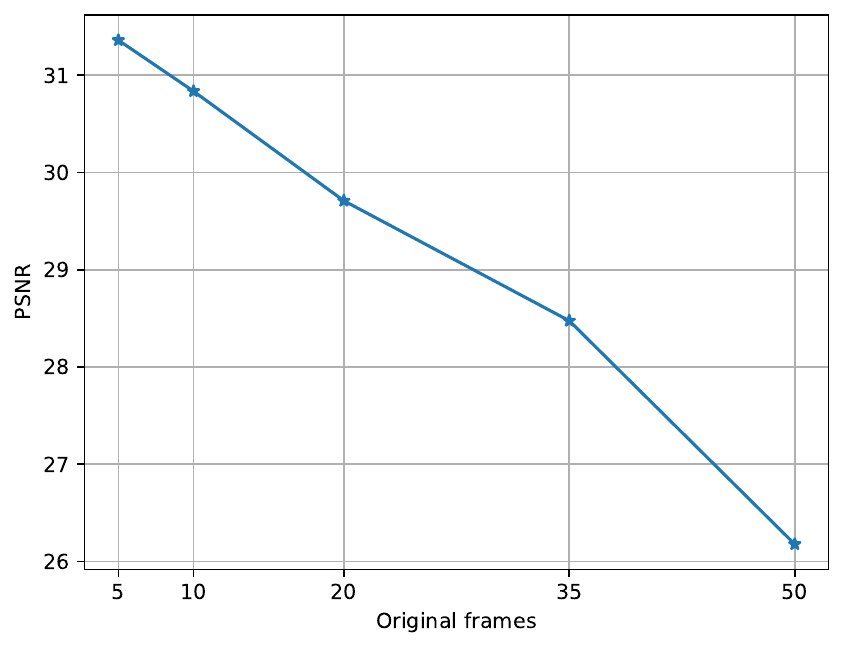}
    \caption{PSNR between reconstruction and ground truth for XCAT-$j$ phantoms, with $j=5, 10, 20, 35, 50$. The motion in XCAT-$j$ increases with $j$, hence, reconstruction quality decreases.}
    \label{fig:psnr frames}
\end{figure}

Figure \ref{fig:rec xcat} shows reconstructions and the corresponding errors against the ground truth image for XCAT-5 and XCAT-50 phantoms at comparable frames that span part of one respiratory cycle. As expected, a larger error is obtained for XCAT-50. This is supported in figure \ref{fig:psnr frames} which shows the PSNR achieved for each phantom after optimization finished and reveals an almost linear decay on the reconstruction with respect to the velocity. We also notice a difficulty in capturing fine details given by the tiny dots (pulmonary alveoli) within the lungs.

The circular structure of the scene with varying sizes is the diaphragm. This structure represents out-of-plane motion, which clearly violates the brightness constancy assumption imposed by the optical flow model. However, given that this is imposed as a soft constraint in the variational problem, our method can still get a reliable reconstruction through the data-fidelity term.

\subsection{Regularized neural fields versus grid-based method}\label{sec:nf vs grid}

\subsubsection{Random sampling}

\begin{figure*}[b]
\centering
\subfloat[Comparing grid-based method against neural fields for the two-square phantom with random sampling. Reconstruction (left) and error (right) using the same regularization parameters $(\alpha,\beta,\gamma)=(10^{-3},10^{-4},10^{-3})$.]{\includegraphics[width=\textwidth]{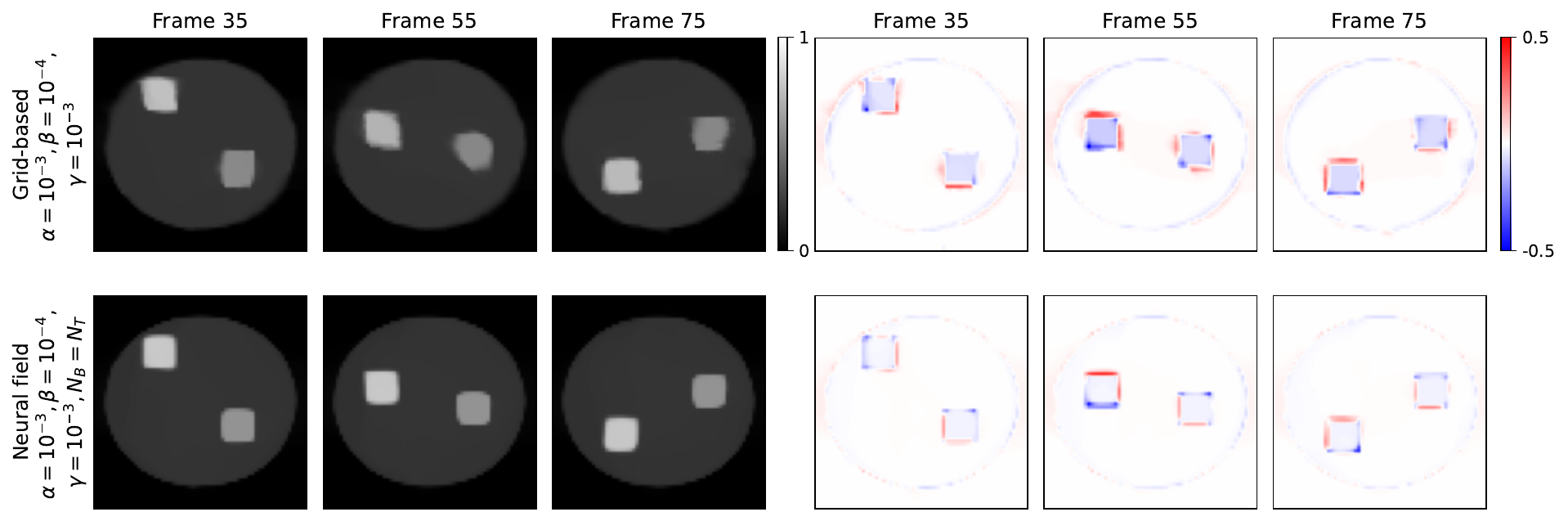}
\label{fig:two-square reconstructions voxel random}}
\vspace{-0.1cm}
\centering
\subfloat[$x-t$ slice view at $y=0$ for the reconstruction of the two-square phantom with random sampling. First column is the ground truth. Second and third columns compare the grid-based and neural field methods with the same regularization parameters. Fourth and fifth columns are the neural field reconstructions from section \ref{sec:unreg vs reg}.]{\includegraphics[width=0.8\textwidth]{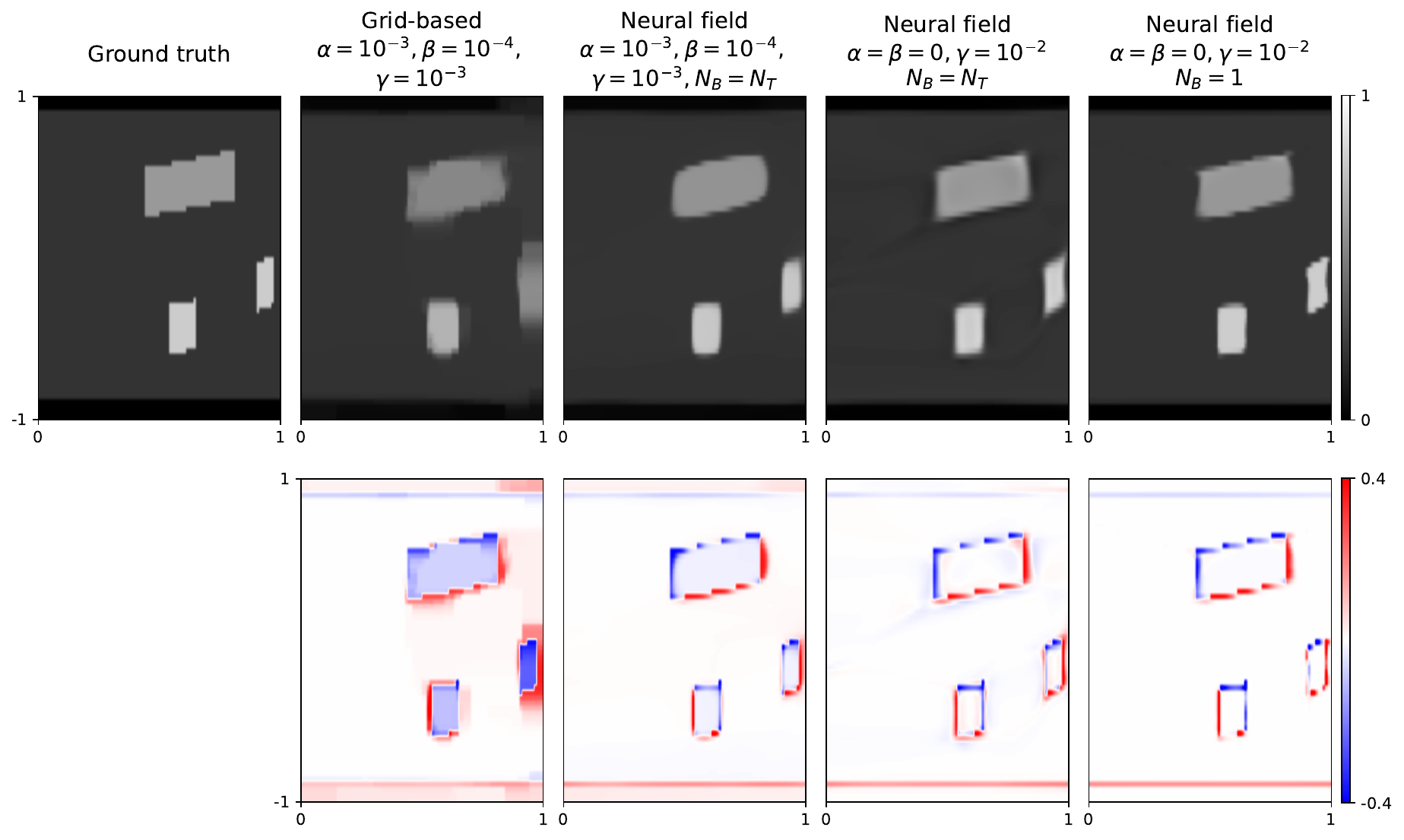}
\label{fig:two-square reconstructions motion slice}}
\caption{Neural field versus grid-based for the two-square phantom. Neural fields achieve smaller errors when compared to the grid-based solution.}
\end{figure*}

\begin{figure*}[]
\centering
\subfloat[Comparing grid-based method against neural fields for the cardiac phantom with random sampling. Reconstruction (left) and error (right) using the same regularization parameters $(\alpha,\beta,\gamma)=(10^{-4}, 10^{-4}, 5\times10^{-3})$]{\includegraphics[width=\textwidth]{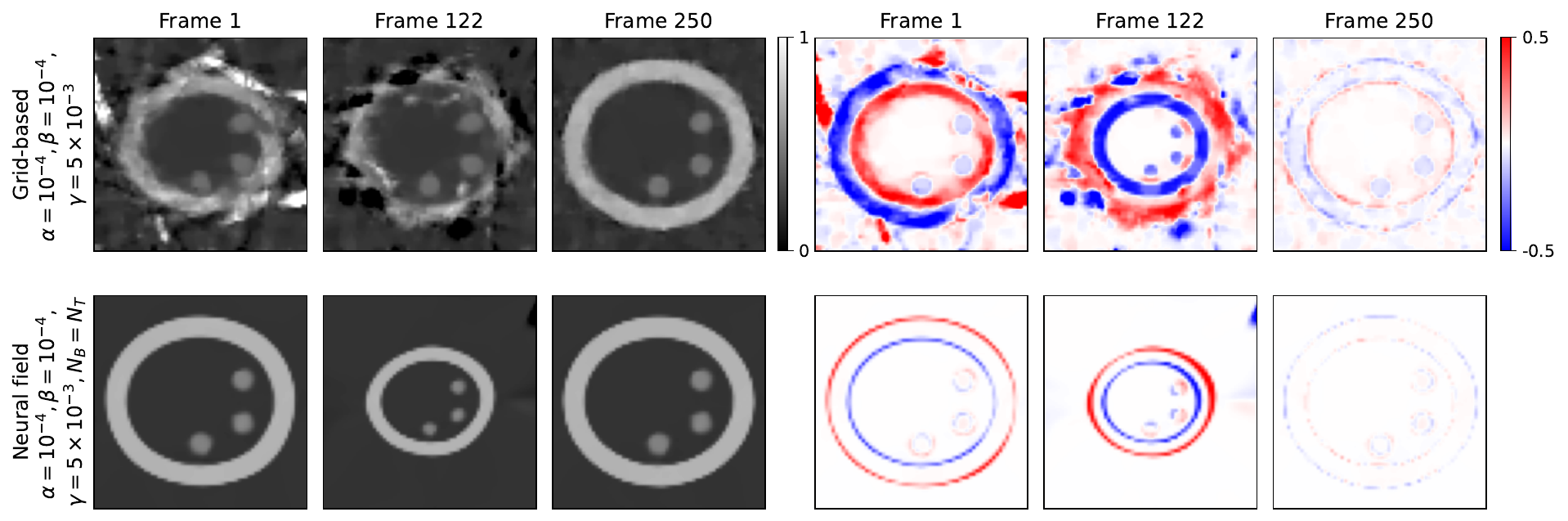}
\label{fig:cardiac reconstructions voxel random}}
\vspace{-0.1cm}
\subfloat[$y-t$ slice view at $x=0$ for the reconstruction of the cardiac phantom with random sampling. First column is the ground truth. Second and third columns compare the grid-based and neural field methods with the same regularization parameters. Fourth and fifth columns are the neural field reconstructions from section \ref{sec:unreg vs reg}.]{\includegraphics[width=0.8\textwidth]{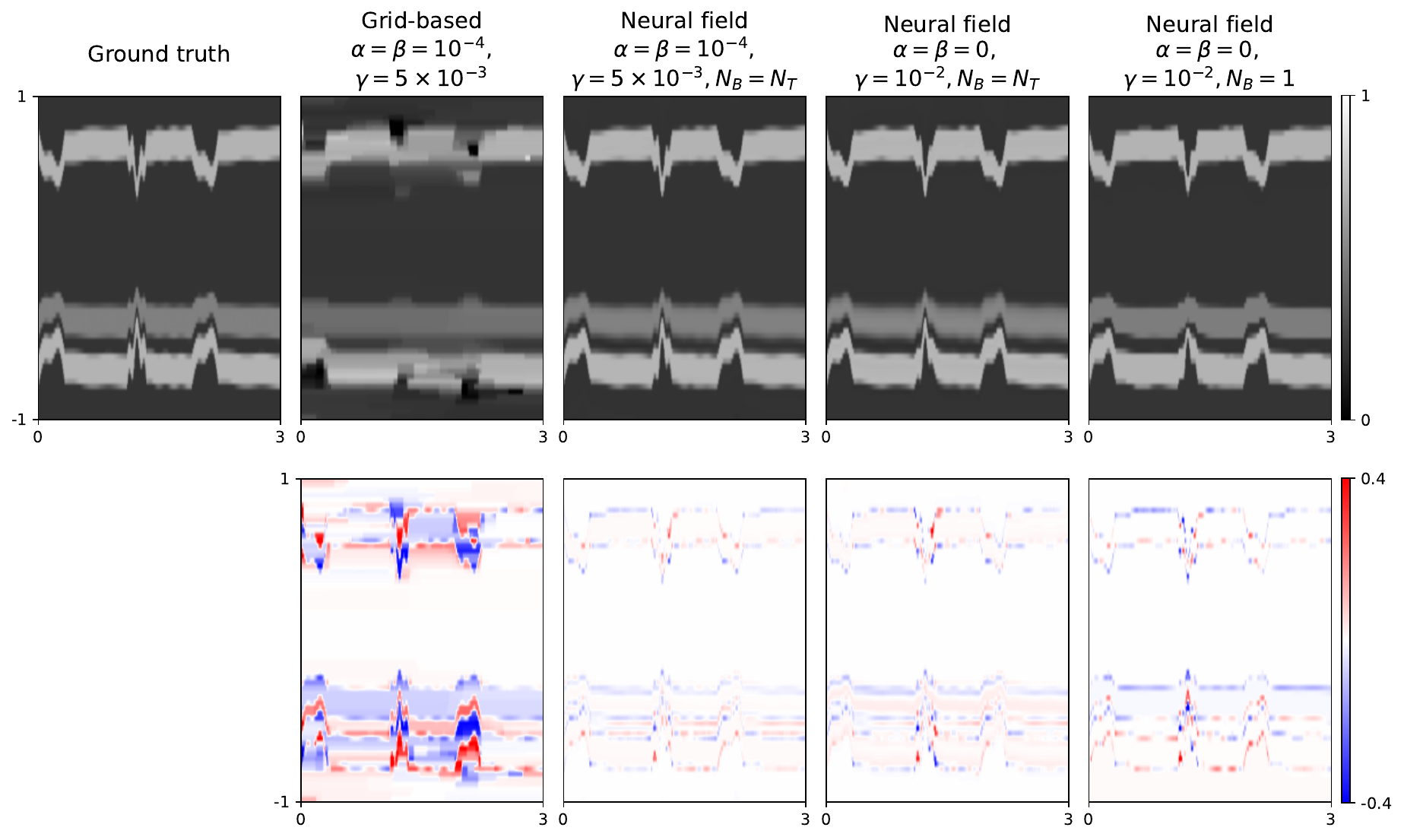}
\label{fig:cardiac reconstructions motion slice}}
\caption{Neural field versus grid-based for the cardiac phantom. The grid-based method fails at the reconstruction of the cardiac phantom while neural fields can capture the intricate motion.}
\end{figure*}

\begin{table*}[b]
\centering
\begin{tabular}{@{}ccccc@{}}
\cmidrule(l){2-5}
 & \multicolumn{2}{c|}{\begin{tabular}[c]{@{}c@{}}Two-square\\ $(\alpha,\beta,\gamma)=(10^{-3}, 10^{-4}, 10^{-3})$\end{tabular}} & \multicolumn{2}{c}{\begin{tabular}[c]{@{}c@{}}Cardiac\\ $(\alpha,\beta,\gamma)=(10^{-4}, 10^{-4}, 5\times10^{-3})$\end{tabular}} \\ \cmidrule(l){2-5} 
\multicolumn{1}{c|}{} & \multicolumn{1}{c|}{Grid-based} & \multicolumn{1}{c|}{Neural Field} & \multicolumn{1}{c|}{Grid-based} & \multicolumn{1}{c|}{Neural Field} \\ \midrule
\multicolumn{1}{|c|}{PSNR} & \multicolumn{1}{c|}{27.09} & \multicolumn{1}{c|}{32.92} & \multicolumn{1}{c|}{17.55} & \multicolumn{1}{c|}{29.77} \\ \midrule
\multicolumn{1}{l}{} & \multicolumn{1}{l}{} & \multicolumn{1}{l}{} & \multicolumn{1}{l}{} & \multicolumn{1}{l}{} \\ \midrule
\multicolumn{1}{|c|}{Data-fidelity} & \multicolumn{1}{c|}{$6.25\times10^{-5}$} & \multicolumn{1}{c|}{$7.41\times10^{-5}$} & \multicolumn{1}{c|}{$1.22\times10^{-4}$} & \multicolumn{1}{c|}{$8.51\times10^{-5}$} \\ \midrule
\multicolumn{1}{|c|}{$\alpha\mathcal{R}(u)$} & \multicolumn{1}{c|}{$5.1\times10^{-4}$} & \multicolumn{1}{c|}{$5.04\times10^{-4}$} & \multicolumn{1}{c|}{$1.81\times 10^{-4}$} & \multicolumn{1}{c|}{$1.25\times10^{-4}$} \\ \midrule
\multicolumn{1}{|c|}{$\beta\mathcal{S}(v)$} & \multicolumn{1}{c|}{$4.6\times10^{-5}$} & \multicolumn{1}{c|}{$1.05\times10^{-6}$} & \multicolumn{1}{c|}{$4.14\times 10^{-4}$} & \multicolumn{1}{c|}{$4.3\times 10^{-5}$} \\ \midrule
\multicolumn{1}{|c|}{$\gamma\mathcal{A}(r(u,v))$} & \multicolumn{1}{c|}{$1.3\times10^{-4}$} & \multicolumn{1}{c|}{$2.23\times10^{-4}$} & \multicolumn{1}{c|}{$1.12\times 10^{-4}$} & \multicolumn{1}{c|}{$5.21\times10^{-3}$} \\ \midrule
\multicolumn{1}{|c|}{Final loss} & \multicolumn{1}{c|}{$7.53\times10^{-4}$} & \multicolumn{1}{c|}{$8.03\times10^{-4}$} & \multicolumn{1}{c|}{$8.29\times10^{-4}$} & \multicolumn{1}{c|}{$5.46\times10^{-3}$} \\ \bottomrule
\end{tabular}
\caption{Comparing the grid-based method against neural fields for the two-square and cardiac phantoms. We show PSNR and the terms from the loss function. In particular, the regularization terms for the neural field are obtained by evaluating it at the whole spatiotemporal grid.}
\label{table:nf vs voxel random}
\end{table*}

\begin{figure*}
\centering
\includegraphics[width=\textwidth]{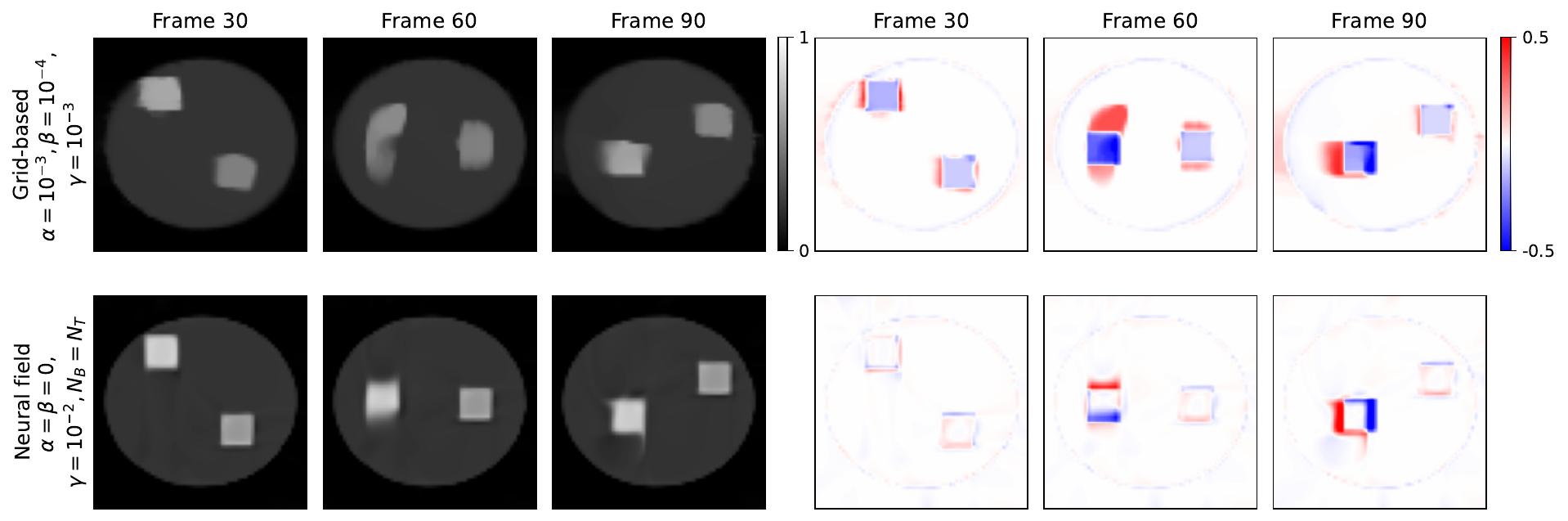}
\caption{Comparing grid-based method against neural fields for the two-square phantom with sequential sampling. Reconstruction (left) and error (right) using the regularization parameters that led to the best results for random sampling. Neural fields still perform better than grid-based but both reconstructions worsen with respect to the one obtained with random sampling.}
\label{fig:two-square reconstructions voxel sequential}
\vspace{0.5cm}
\includegraphics[width=\textwidth]{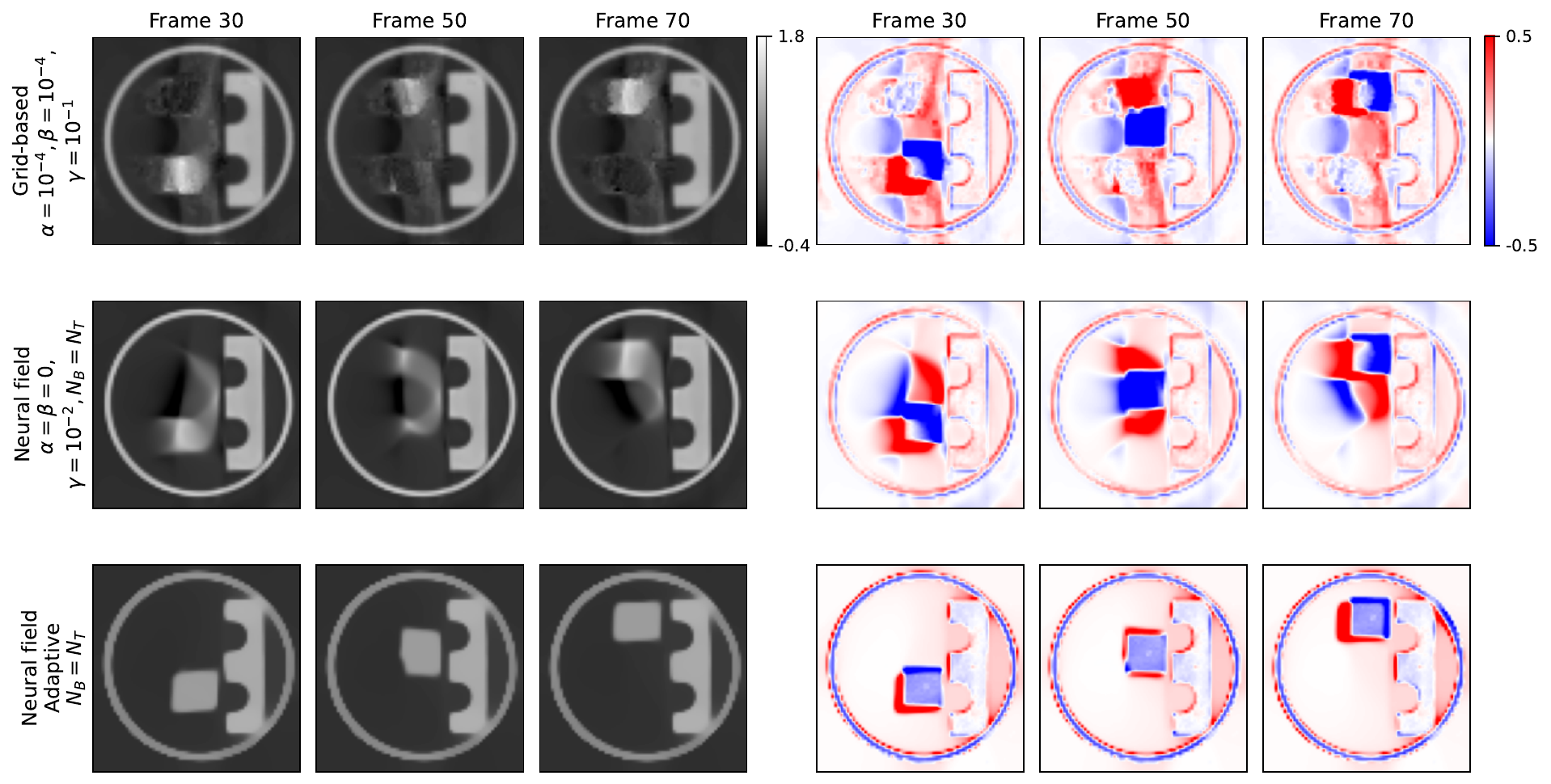}
\caption{Comparing grid-based method against neural fields for the STEMPO phantom with sequential 4-degree sampling. Reconstruction (left) and error (right). Both the neural field and the grid-based solutions fail at the reconstruction task. Neural field Adaptive increases the value of $\gamma$ to improve the reconstruction.}
\label{fig:stempo reconstructions voxel sequential}
 
\end{figure*}

We now compare explicitly regularized neural fields against the grid-based method outlined in section \ref{sec:grid-based method} for the random sampling regime. For this approach, it was found that the choice $(\alpha, \beta, \gamma) = (10^{-3}, 10^{-4}, 10^{-3})$ led to the best results in terms of PSNR, achieving a value of 27.09. We solve the same variational problem with neural fields by choosing the same regularization parameters, in which case the PSNR achieved is 32.92. Figure \ref{fig:two-square reconstructions voxel random} shows reconstructions at different frames. To appreciate the time evolution we also show a $x-t$ slice view in figure \ref{fig:two-square reconstructions motion slice} at the horizontal center $y=0$, where it becomes clear that the grid-based solution struggles at capturing regularity in time. 

The same experiment is performed for the cardiac dataset, for which the best reconstruction was attained at $(\alpha, \beta, \gamma)=(10^{-4}, 10^{-4},5\times10^{-3})$ achieving a PSNR of 17.55 for the grid-based method and 29.77 for the neural field (notice that this value is higher than the PSNR attained for $\gamma=10^{-2}$ in the previous section). In this case, the grid-based method completely fails at the reconstruction task and the rapid motion of this phantom cannot be captured. Figure \ref{fig:cardiac reconstructions motion slice} shows a $y-t$ slice at the vertical center $x=0$ and it can be seen that the inner circle at the bottom barely moves. The neural field on the other hand is still able to show regularity in time. This shows that neural fields can outperform the grid-based solution even for the choice of regularization parameters that led to the best behavior for the grid-based method. 

We report the PSNR and loss values for both experiments in table \ref{table:nf vs voxel random}. For the neural field, the regularizers are estimated by evaluating it on the same cartesian grid as for the grid-based representation. There we can see that the grid-based achieves a lower loss for both experiments, in particular, it attains a lower value for the data fidelity and the optical flow terms. We highlight that a very low optical flow is related to static motion as observed in the cardiac phantom in figure \ref{fig:cardiac reconstructions motion slice}.

\subsubsection{Sequential sampling}

We finish our study by comparing both methods at the more challenging problem of sequential sampling. We try the sequential 9-degree sampling for the two-square phantom and the 4-degree sampling for STEMPO. For the two-square phantom, we use the regularization parameters that gave the best results in the previous section, namely, $(\alpha, \beta, \gamma)=(10^{-3}, 10^{-4}, 10^{-3})$ for the grid-based method achieving a PSNR of 22.65, while for the neural field, we use
$(\alpha,\beta,\gamma)=(0,0,10^{-2})$ and the method achieves a PSNR of 26.42.  Results are shown in figure \ref{fig:two-square reconstructions voxel sequential}. There it can be seen that both methods perform worse than in the random sampling, but still, the neural field reconstruction is better than the grid-based one which shows large errors, for instance, in frame 60. For STEMPO we found $(\alpha,\beta, \gamma)=(10^{-4},10^{-4},10^{-1})$ to give the best reconstruction for the grid-based method with a PSNR of 14.24; for the neural field, we set $(\alpha,\beta,\gamma)=(0,0,10^{-2})$, leading to a PSNR of 15.18. Results can be seen in figure \ref{fig:stempo reconstructions voxel sequential}. In this case, both methods can capture the static part of the image but fail at the reconstruction of the moving square due to the sampling scheme. 

The bad reconstructions shown in the first two rows of figure \ref{fig:stempo reconstructions voxel sequential} motivated us to try different strategies for the neural field. In particular, we observe a different behavior for the choice $(\alpha,\beta,\gamma)=(10^{-3}, 10^{-4}, 10^{-3})$: during optimization, the neural field fits the static part but after 15,000 iterations approximately, the optical flow error begins to increase, promoting a better reconstruction of the dynamic part but, at the same time, worsening the static part. This behavior can be captured during the optimization since we have access to the optical flow loss. Thus, we try an adaptive routine, where the value of $\gamma$ is increased if the optical flow error increases as well. We increase the value of $\gamma$ from $10^{-3}$ to $10^{-2}$ when the optical flow error increases. This reconstruction is shown in the third row of figure \ref{fig:stempo reconstructions voxel sequential}. The PSNR achieved is now 17.39 and both static and dynamic parts are better captured. This shows that neural fields can potentially solve the challenging sequential sampling case with more dedicated optimization routines. We leave this for future research. We do not try such a strategy for the grid-based case, hence, we cannot conclude that this is a particular benefit of the neural field representation.

\newpage
\section{Discussion}

Despite neural fields being regarded as resolution-independent representations, they still need to be queried at spatiotemporal grid coordinates to evaluate the forward operator. Since the evaluation of each coordinate requires a forward pass of the network, this process may become computationally inefficient and time-consuming for large-scale problems (see, for instance, the runtimes in figure~\ref{fig:batch size}). Thus, a limitation of the spatiotemporal neural field used in this work arises when scaling to a 3D+time scenario with demand for high spatial resolution, as in dynamic cone-beam CT.

This limitation can be addressed by novel positional encodings, such as hash enconding \cite{muller2022instant} or ACORN \cite{martel2021acorn}. These encodings work at multiscale resolution levels and can capture fine details at a lower computational cost. In particular, they resulted in faster and higher-quality reconstructions compared to the Fourier encoding used in this work. Another line of research accelerates the optimization by applying the forward operator on auxiliary image variables instead of acting on the rasterized image obtained from the neural field at every iteration \cite{lozenski2024proxnf, najaf2025accelerated}. In such approaches, weight updates occur entirely in the image domain. Importantly, the PDE-based regularizer employed in this work can be seamlessly integrated into such frameworks, highlighting its potential when scaling to more realistic and computationally demanding settings.

\section{Conclusion}\label{sec:conclusions}

This work considers neural fields for dynamic CT image reconstruction from finely resolved in time but severely undersampled in space measurements as commonly encountered in applications, e.g. cardiac CT. Neural fields are particularly suitable for this task: their continuity allows for a coherent representation in both time and space; their resolution independence enables us to query it at multiple frames, and, together with their differentiability, facilitate the numerical evaluation of the PDE-based motion regularizer. We studied how to enhance the neural field reconstruction by making use of the optical flow equation: constraining the neural field to this physically feasible motion meant a significant improvement with respect to the state-of-the-art both in terms of neural fields as well as traditional grid-based representations. We finish by mentioning that further regularization techniques for dynamic imaging with neural fields can be studied, such as sparsity in suitable domains or deep-learning-based ones. This is left for future research.

\bmhead{Acknowledgments}
PA is supported by a scholarship from the EPSRC Centre for Doctoral Training in Statistical Applied Mathematics at Bath (SAMBa), under the project EP/S022945/1. MJE acknowledges support from the EPSRC (EP/S026045/1, EP/T026693/1, EP/V026259/1, EP/Y037286/1) and the European Union Horizon 2020 research and innovation programme under the Marie Skodowska-Curie grant agreement REMODEL. The authors gratefully acknowledge the University of Bath’s Research Computing Group (doi.org/10.15125/b6cd-s854) for their support in this work.

\bibliography{sn-bibliography}

\end{document}